\documentstyle[galley,epsf]{mn}

\newif\ifAMStwofonts

  \newcommand{\Teff}{\mbox{\,\em T$_{\rm eff}$}}         
  
  \newcommand{\ion}[2]{\mbox{\,#1\,{\sc #2}}}         
  \newcommand{\rcb}{R~CrB\ }
  \def\simge{\mathrel{\raise1.16pt\hbox{$>$}\kern-7.0pt
    \lower3.06pt\hbox{{$\scriptstyle \sim$}}}}           
  \def\simle{\mathrel{\raise1.16pt\hbox{$<$}\kern-7.0pt
    \lower3.06pt\hbox{{$\scriptstyle \sim$}}}}           
\ifoldfss
  \ifCUPmtlplainloaded \else
    \NewTextAlphabet{textbfit} {cmbxti10} {}
    \NewTextAlphabet{textbfss} {cmssbx10} {}
    \NewMathAlphabet{mathbfit} {cmbxti10} {} 
    \NewMathAlphabet{mathbfss} {cmssbx10} {} 
  \fi
  \ifAMStwofonts
    \ifCUPmtlplainloaded \else
      \NewSymbolFont{upmath} {eurm10}
      \NewSymbolFont{AMSa} {msam10}
      \NewMathSymbol{\upi}     {0}{upmath}{19}
      \NewMathSymbol{\umu}     {0}{upmath}{16}
      \NewMathSymbol{\upartial}{0}{upmath}{40}
      \NewMathSymbol{\leqslant}{3}{AMSa}{36}
      \NewMathSymbol{\geqslant}{3}{AMSa}{3E}

      \let\leq=\leqslant 
      \let\geq=\geqslant 
    \fi
  \fi
\fi 

\ifnfssone
  \newmathalphabet{\mathit}
  \addtoversion{normal}{\mathit}{cmr}{m}{it}
  \addtoversion{bold}{\mathit}{cmr}{bx}{it}
  \newmathalphabet{\mathbfit} 
  \addtoversion{normal}{\mathbfit}{cmr}{bx}{it}
  \addtoversion{bold}{\mathbfit}{cmr}{bx}{it}
  \newmathalphabet{\mathbfss} 
  \addtoversion{normal}{\mathbfss}{cmss}{bx}{n}
  \addtoversion{bold}{\mathbfss}{cmss}{bx}{n}
  \ifAMStwofonts
    \ifCUPmtlplainloaded \else
      \UseAMStwoboldmath
      \makeatletter
      \new@mathgroup\upmath@group
      \define@mathgroup\mv@normal\upmath@group{eur}{m}{n}
      \define@mathgroup\mv@bold\upmath@group{eur}{b}{n}
      \edef\UPM{\hexnumber\upmath@group}
      \new@mathgroup\amsa@group
      \define@mathgroup\mv@normal\amsa@group{msa}{m}{n}
      \define@mathgroup\mv@bold\amsa@group{msa}{m}{n}
      \edef\AMSa{\hexnumber\amsa@group}
      \makeatother
      \mathchardef\upi="0\UPM19
      \mathchardef\umu="0\UPM16
      \mathchardef\upartial="0\UPM40
      \mathchardef\leqslant="3\AMSa36
      \mathchardef\geqslant="3\AMSa3E

      \let\leq=\leqslant 
      \let\geq=\geqslant 
    \fi
  \fi
\fi 

\ifnfsstwo
  \DeclareMathAlphabet{\mathbfit}{OT1}{cmr}{bx}{it}
  \SetMathAlphabet\mathbfit{bold}{OT1}{cmr}{bx}{it}
  \DeclareMathAlphabet{\mathbfss}{OT1}{cmss}{bx}{n}
  \SetMathAlphabet\mathbfss{bold}{OT1}{cmss}{bx}{n}
  \ifAMStwofonts
    \ifCUPmtlplainloaded \else
      \DeclareSymbolFont{UPM}{U}{eur}{m}{n}
      \SetSymbolFont{UPM}{bold}{U}{eur}{b}{n}
      \DeclareSymbolFont{AMSa}{U}{msa}{m}{n}
      \DeclareMathSymbol{\upi}{0}{UPM}{"19}
      \DeclareMathSymbol{\umu}{0}{UPM}{"16}
      \DeclareMathSymbol{\upartial}{0}{UPM}{"40}
      \DeclareMathSymbol{\leqslant}{3}{AMSa}{"36}
      \DeclareMathSymbol{\geqslant}{3}{AMSa}{"3E}

      \let\leq=\leqslant 
      \let\geq=\geqslant 
    \fi
  \fi
\fi 

\ifCUPmtlplainloaded \else
  \ifAMStwofonts \else 
    \def\upi{\pi}
    \def\umu{\mu}
    \def\upartial{\partial}
  \fi
\fi

\title{Abundance analyses of cool extreme helium stars}
\author[Gajendra Pandey et al.]
       {Gajendra Pandey,$^1$ N. Kameswara Rao,$^2$ David L. Lambert,$^1$ 
        C. Simon Jeffery,$^3$ 
\newauthor
Martin Asplund $^4$\\
       $^1$Department of Astronomy, University of Texas, Austin, TX 78712-1083, USA\\
       $^2$Indian Institute of Astrophysics, Bangalore 560034, India\\
        $^3$Armagh Observatory, College Hill, Armagh BT61 9DG, Northern Ireland\\
        $^4$Astronomiska Observatoriet, Box 515, S-751 20 Uppsala, Sweden\\}
    
\date{Accepted .
      Received ;
      in original form 2000 }

\pagerange{\pageref{firstpage}--\pageref{lastpage}}
\pubyear{2000}

\begin{document}

\maketitle

\label{firstpage}
\begin{abstract}

Extreme helium stars (EHe) with effective temperatures from 8000K to 13000K
 are among the coolest EHe stars and overlap the hotter \rcb stars in
effective temperature. The cool EHes
may represent an evolutionary link between the hot
EHes and the R~CrBs. 
Abundance analyses of four cool EHes,
BD$ +1^\circ 4381$ (FQ Aqr), LS IV --14$^\circ$ 109, BD --1$^\circ$ 3438 (NO Ser),
and LS IV $-1^\circ 002$ (V2244 Oph) are presented.
All these stars show evidence of H- and He-burning at earlier stages of their
evolution. \\

To test for an evolutionary connection, the chemical compositions
of cool EHes are compared with those of hot EHes
and R~CrBs. Relative to Fe, the N abundance of these stars is intermediate between those
of hot EHes and R~CrBs.
For the R~CrBs, the metallicity M derived
from the mean of Si and S appears to be more consistent with the kinematics than that
derived from Fe.
When metallicity M derived from Si and S replaces Fe, the observed N abundances
of EHes and R~CrBs fall at or below the upper limit corresponding to thorough
conversion of initial C and O to N. 
There is an apparent difference between the composition
of R~CrBs and EHes; the former having systematically higher [N/M] ratios.
The material present in the atmospheres of many R~CrBs is heavily 
CN- and ON-cycled. Most of the EHes have only CN-cycled 
material in their atmospheres. There is an indication that
the CN- and ON-cycled N in EHes was partially converted to Ne by $\alpha$-captures.
If EHes are to evolve to R~CrBs, fresh C in EHes has to be converted
to N; the atmospheres of EHes have just sufficient hydrogen to raise the 
N abundance to the level of R~CrBs. If Ne is found to be normal in R~CrBs,
the proposal that EHes evolve to R~CrBs fails.
The idea that R~CrBs evolve to EHes is ruled out; the N abundance in R~CrBs has to be reduced to the level of EHes,
as the C/He which is observed to be uniform across EHes, has to be maintained.
Hence, the inferred [N/M], C/He, [Ne/M], and the H-abundances of these two
groups indicate that the EHes and the R~CrBs may not be on the same evolutionary path.\\

The atmospheres of H-deficient stars most likely consist of three ingredients:
a residue of normal H-rich material, substantial
amounts of H-poor CN(O)-cycled material, and C- (and O-) rich
material from gas exposed to He-burning. This composition could be a result of final He-shell
flash in a single post-AGB star (FF scenario), or a merger of two white dwarfs (DD scenario).
Although the FF scenario
accounts for Sakurai's object and other stars
(e.g., the H-poor central stars of planetary nebulae),
present theoretical calculations imply higher
C/He and O/He ratios than are observed in EHes and R~CrBs.
Quantitative predictions are lacking for the DD scenario.

\end{abstract}


\section{Introduction}

The extreme helium stars (hereafter, EHe) are a rare class of
stars. Popper (1942) discovered the first EHe, HD\,124448, and
Thackeray \& Wesselink (1952) the second, HD\,168476. Today, about 25
are known (Jeffery 1996). Early work on the surface composition
of the EHes by curve-of-growth techniques, notably by Hill (1964, 1965),
concentrated on the hotter EHes whose spectra are characterized by strong
lines of neutral helium, singly ionized carbon, and weak or
absent Balmer lines.
The first self-consistent spectroscopic analysis
 using hydrogen-deficient model atmospheres was
performed by Sch\"{o}nberner \& Wolf (1974) for Popper's star.
Jeffery (1996) reviews  modern work on abundance analyses
of EHes. More recent work includes the spectral analyses of
HD 144941 (Harrison \& Jeffery 1997; Jeffery \& Harrison 1997), LSS 3184 (Drilling, Jeffery, \& Heber 1998),
LS IV +6$^\circ$ 002 (Jeffery 1998), LSS 4357, LS II +33$^\circ$005 and
LSS 99 (Jeffery et al. 1998) and V652 Her (Jeffery, Hill, \& Heber 1999). Published
work has emphasized the hotter EHes. Cooler EHes, stars with
effective temperatures of 8000K to 13000K, have been largely ignored.
A quartet of such EHes is analysed here.

Our stars were selected from
the list provided by Jeffery et al. (1996).
  Three of our stars $-$ BD +1$^\circ$ 4381 (FQ Aqr), LS IV $-14^\circ$ 109
 and LS IV $-1^\circ$ 002 (NO Ser)
$-$ were discovered by Drilling during the course
 of a spectroscopic survey of stars
down to photographic magnitude 12.0, which
 had been classified as OB$^+$ stars in the Case-Hamburg
 surveys (Drilling 1979, 1980). The final member of the quartet,
BD --1$^\circ$ 3438 (V2244 Oph), is one of  eight
extreme helium stars described by Hunger (1975).
Judged by effective temperature and luminosity, these cool EHes may represent
an evolutionary link between the hot EHes and the R~CrBs. A major goal of
our abundance analyses was to test this link using chemical
compositions of the three groups of stars. Our abundance analyses
are based on high-resolution optical spectra
and model atmospheres.\footnote{Preliminary analyses of
FQ\,Aqr and LS IV --1$^\circ$\,002 are reported by Asplund et al. (2000).}

\section{Observations}
 
High-resolution optical spectra of the four EHe stars
were obtained on 25 July 1996 at the W. J. McDonald Observatory 2.7-m
telescope with the
coud\'{e} cross-dispersed echelle
spectrograph (Tull et al. 1995) at a 2-pixel
resolving power (R =$\lambda/\Delta\lambda$) of 60,000.
The detector was a Tektronix 2048 $\times$ 2048 CCD. The recorded spectrum
covered the wavelength range from 3800\AA\ to 10000\AA,
but the spectral coverage was incomplete longward of
about 5500\AA. A Th-Ar hollow cathode lamp
 was observed either just prior
to or just after exposures of the programme stars to
 provide wavelength calibration.
In order to remove the pixel-to-pixel variation
 in the sensitivity of the CCD,
exposures were obtained of
a halogen lamp. Typical exposure
times of our programme stars were 30 minutes, and two exposures were co-added
to improve the signal-to-noise ratio of the final spectrum,
and to identify and eliminate cosmic rays.
To cover the missing wavelength regions in the red,
observations were made at a slightly different
grating setting on 26 July 1996 for two
of the programme stars, FQ Aqr and BD --1$^\circ$ 3438.
The FWHM of the Th-Ar comparison lines and the atmospheric lines present in the spectra corresponds to
6.0 km s$^{-1}$. We have used the Image Reduction and Analysis Facility (IRAF) software
packages to reduce the spectra.


Lines  were identified using the  Revised Multiplet Table (Moore 1972),
the selected Tables of Atomic Spectra (Moore 1970), the
line list provided by Kurucz \& Peytreman (1975),
and also  the investigations of Hill (1964, 1965), Lynas-Gray et al. (1981) and Heber (1983).
Lines of  \ion{C}{i}, \ion{C}{ii} and \ion{He}{i} lines were readily identified (Figure 1).
No lines of \ion{He}{ii} were found.
Lines of all elements
expected and observed in early A-type and late B-type normal stars  were found.
Lines of ionized metals of the iron group are plentiful.
These lines are much stronger when compared with those observed in early A-type and late B-type
normal stars, a notable
feature of the spectra of cool EHe stars and  attributable to the  lower
opacity in the atmosphere due to hydrogen deficiency.
Our large spectral coverage enabled us to identify several important elements
in one or two stages of ionization (see Figure 1).

\begin{figure}
\epsfxsize=8truecm
\epsffile{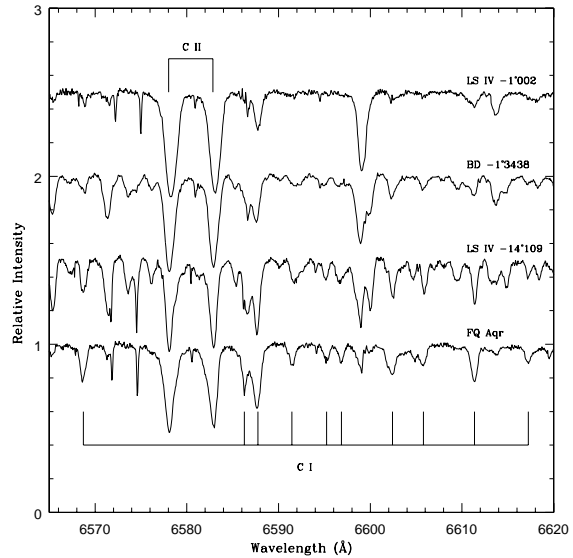}
\caption{Region showing neutral and ionized lines of \ion{C}{i} and \ion{C}{ii}
in the spectra of cool EHe stars}
\end{figure}

The photospheric radial velocities (R.V.) for the programme stars were measured using 
\ion{Fe}{ii} lines and are listed in Table 1. The number of \ion{Fe}{ii} lines used 
is given within brackets.

\section{Model Atmospheres of hydrogen-deficient stars}

Our analyses are based on model atmospheres constructed from the classical
assumptions:
 the energy (radiation plus convection)
 flux is constant, and  the atmosphere consists of
plane-parallel layers in hydrostatic and local thermodynamic equilibrium (LTE).
For the abundance analysis, a model was combined with the
appropriate  programme to predict equivalent widths and, by iteration,
to obtain the elemental abundances.

Analyses have to be self-consistent, i.e., the derived abundances should
be identical to those used in constructing the model. For the cooler stars,
the key quantity
is the C/He ratio (Asplund et al. 1997a). Available model grids cover an adequate
range in the C/He ratio. For the hot stars, helium is a controlling
influence on the atmospheric structure as such the C/He ratio plays
a minor role.

\subsection{Models for \Teff~$\leq$ 9,500~K}

For stars with \Teff~$\leq$ 9,500 K, we used the Uppsala line-blanketed
models described by Asplund et al. (1997a).
The important features of this model  grid are the inclusion of
line-blanketing by opacity sampling,
 and the use of modern values for continuous opacities
 from the
Opacity project (Seaton et al. 1994 and references therein). Free-free opacity
of \ion{He}{i}, \ion{C}{i} and \ion{C}{ii} (Peach 1970) have been incorporated.
Electron and Rayleigh scattering are included.

The  grid provides  models
 for temperatures in the range 5000 K $\leq$ \Teff~$\leq$ 9500 K
and gravities in the range $-$0.5 $\leq$ log {\it g} $\leq$ 2.0 [cgs].
The abundances used for the standard grid are mainly taken from
Lambert \& Rao (1994). Models are calculated for the following values
of the C/He ratio: C/He = 0.1, 0.3, 1.0, 3.0, and 10.0\% with other elements at fixed
values. That the derived abundances of elements other than C and He
may differ slightly from their assumed values is not likely to be a serious
source of error.
The line formation calculations were carried out with the Uppsala LTE
line-formation code EQWIDTH.

\subsection{Models for \Teff~$\geq$ 10,000~K}

To analyse cool EHe stars in the temperature range
10000 K $\leq$ \Teff~$\leq$ 14000 K, a grid of appropriate models was
calculated using the model atmosphere code STERNE (Jeffery \& Heber 1992).
The grid provides models for
temperatures in the range 10000 K $\leq$ \Teff~$\leq$ 40000 K, and
gravities in the range 1.0 $\leq$ log {\it g} $\leq$ 8.0 [cgs].
The relative abundances by number used for the standard grid are 
He = 99\%, H/He = 10$^{-4}$, C/He = 1\% and
the rest of the elements are solar. Models
were computed in the
temperature range 10000 K $\leq$ \Teff~$\leq$ 14000 K, for the following
values of C/He ratios: C/He = 0.1\%, 0.3\%, 0.5\%, 1.0\%, 3.0\%, and 10.0\%,
with other elements at fixed values.
The opacity calculations
were made after taking into account the effects of line-blanketing using the
tables of opacity distribution functions for helium- and carbon-rich material.
 The Belfast LTE code SPECTRUM was used for line formation calculations
(Jeffery \& Heber 1992; Jeffery, unpublished).

\subsection{Consistency between model grids}

The two model atmosphere grids
do not overlap in effective temperature.
To compare the grids, we derived a model for 9500K by extrapolating
the high temperature grid models whose coolest models are
at 10000K. The extrapolated model and an Uppsala model for
9500K gave identical abundances to within 0.05 dex.
 The abundances derived for weak lines using SPECTRUM and EQWIDTH
    are in agreement within 0.1 dex for most of the species. We find that
    the abundances derived using the former are always lower than those derived
    using the latter. This small difference is probably due
    to the data used for continuous opacity being from two different sources.

\subsection{Abundance Analysis - Some fundamentals}

Attention
has to be paid to both the line and the continuous opacities in
extracting the abundance of an element $E$ from lines produced by an atom, ion,
or molecule of $E$. In the case of normal stars, hydrogen
directly or indirectly exerts a major influence on the
continuous opacity with the result that analysis of lines of element
 $E$ provides the abundance $E/H$ without recourse to a direct
measurement of the hydrogen abundance from \ion{H}{i} lines. The result is
dependent on the  assumed He/H ratio which is small ($\simeq 0.1$) for
normal stars and unlikely to vary greatly from one normal star
to the next.
Since $E$ is a minor species and He/H is effectively common
to all normal stars, $A_E=E/H$ is a fair measure of abundance. Some
authors prefer to quote abundance as a mass fraction, say $Z(E)$
where

\begin{eqnarray}
Z(E) &=& \frac{\mu_EN_E}{\mu_HN_H + \mu_{He}N_{He}\, ... +\mu_iN_i}  \nonumber \\
     & \simeq& \frac{\mu_EA_E}{1 + 4A_{He}}
\end{eqnarray}

where $Z(E)$ is directly calculable from $A_E$, the fruits of the
abundance analysis, and an assumption about $A_{He}$.
 Of course,
this latter assumption may be replaced by a spectroscopic
measurement in the case of hot stars whose spectra provide helium lines.

Elemental mass fraction is an invaluable quantity when the hydrogen to
helium abundance has been changed by the addition of nuclear-processed
material from H and He burning layers, and comparisons are to be
made between normal and peculiar stars.
The number of nucleons is conserved.
Changes to the hydrogen and helium abundance are unlikely
to be accompanied by a change of (say)
 the iron content of the atmosphere and, hence, the
mass fraction of iron will be unaltered even though hydrogen may have
been greatly depleted.
In the case of cool He-rich stars like the R~CrBs, carbon is the
source of the continuous opacity so that analysis of lines of $E$ gives the
ratio $E/C$.
 The mass fraction $Z(E)$ for the H-poor case is given by

\begin{equation}
Z(E) = \frac{\mu_EA^{\prime}_E}{A^{\prime}_H + 4{\rm He/C} + 12\,  ... + \mu_iA^{\prime}_i}
\end{equation}

where $A^\prime_E = E/C$. In recognition of the  conservation of nucleons, it
is helpful to normalize the customary abundances based on the convention
that log$\epsilon(E) = log(E/H) + 12.0$ to a scale in which
log$\Sigma\mu_i\epsilon(i) = 12.15$ where the constant of 12.15 is derived
from the solar abundances with He/H $\simeq 0.1$;
we write these normalized abundances as log$\epsilon(E)$.
 On this scale, if all hydrogen is converted to helium, the helium
abundance is about 11.54. For approximately solar abundances,
elements carbon and heavier
contribute 0.01 dex or less to the sum.

Since the He/C ratio is most probably large, say about 100,
 $Z(E) \simeq \mu_EA_E^\prime/(4{\rm He/C})$ which is not
calculable from the $E/C$ without  
either an assumption
about or a measurement of the He/C.
The ratio is not spectroscopically determinable for
cool H-poor stars and, there are not,
unlike the He/H ratio
of normal stars, astrophysical grounds for asserting that the C/He
ratio is likely to have a particular value.
Stellar kinematics can provide a guide for initial metallicity and in turn constrain
the He/C (Rao \& Lambert 1996).
 Of course, abundance ratios of elements
$E_1$ and $E_2$ are not dependent on the unknown C/He except in so far as
the model atmosphere structure is dependent on C/He.

To illustrate when the C/He ratio is or
 is not directly determinable from
the spectrum of a H-poor star, we present the results of
two  calculations: an investigation of the sources of
continuous opacity and predictions of the
 equivalent widths of representative lines of \ion{C}{i}, \ion{C}{ii}, and \ion{He}{i}
for model atmospheres spanning the  effective temperature range of
interest.

Figures 2, 3 and 4 show the run of continuous opacity at 5000 \AA~for
the major contributors as a
      function of optical depth
for the following model atmospheres:\\

\indent
      (i) \Teff~= 9500 K, $\log g$ = 1.0 and C/He = 1\%,\\
\indent
      (ii) \Teff~= 11000 K, $\log g$ = 1.0 and C/He = 1\%, and \\
\indent
      (iii) \Teff~= 13000 K, $\log g$ = 2.0 and C/He = 1\%.\\

\begin{figure}
\epsfxsize=8truecm
\epsffile{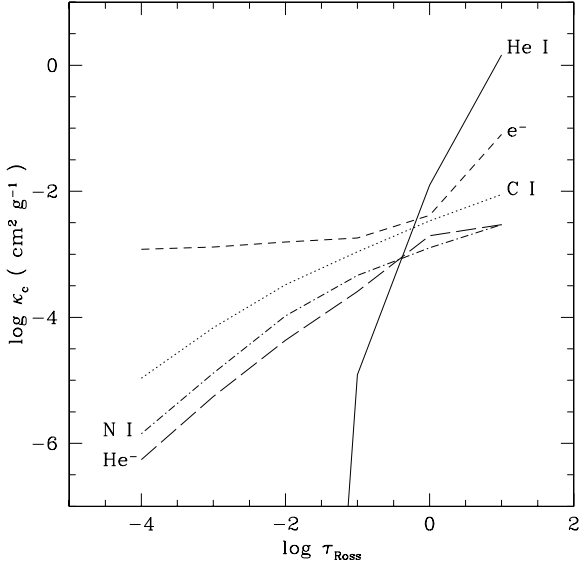}
\caption{Dominant sources of continuous opacity at 5000 \AA~($\kappa_c$) as a function
of Rosseland mean optical depth ($\tau_{Ross}$) for the model atmosphere:
\Teff~= 9500 K, $\log g$ = 1.0 and C/He = 1\%.}
\end{figure}

\begin{figure}
\epsfxsize=8truecm
\epsffile{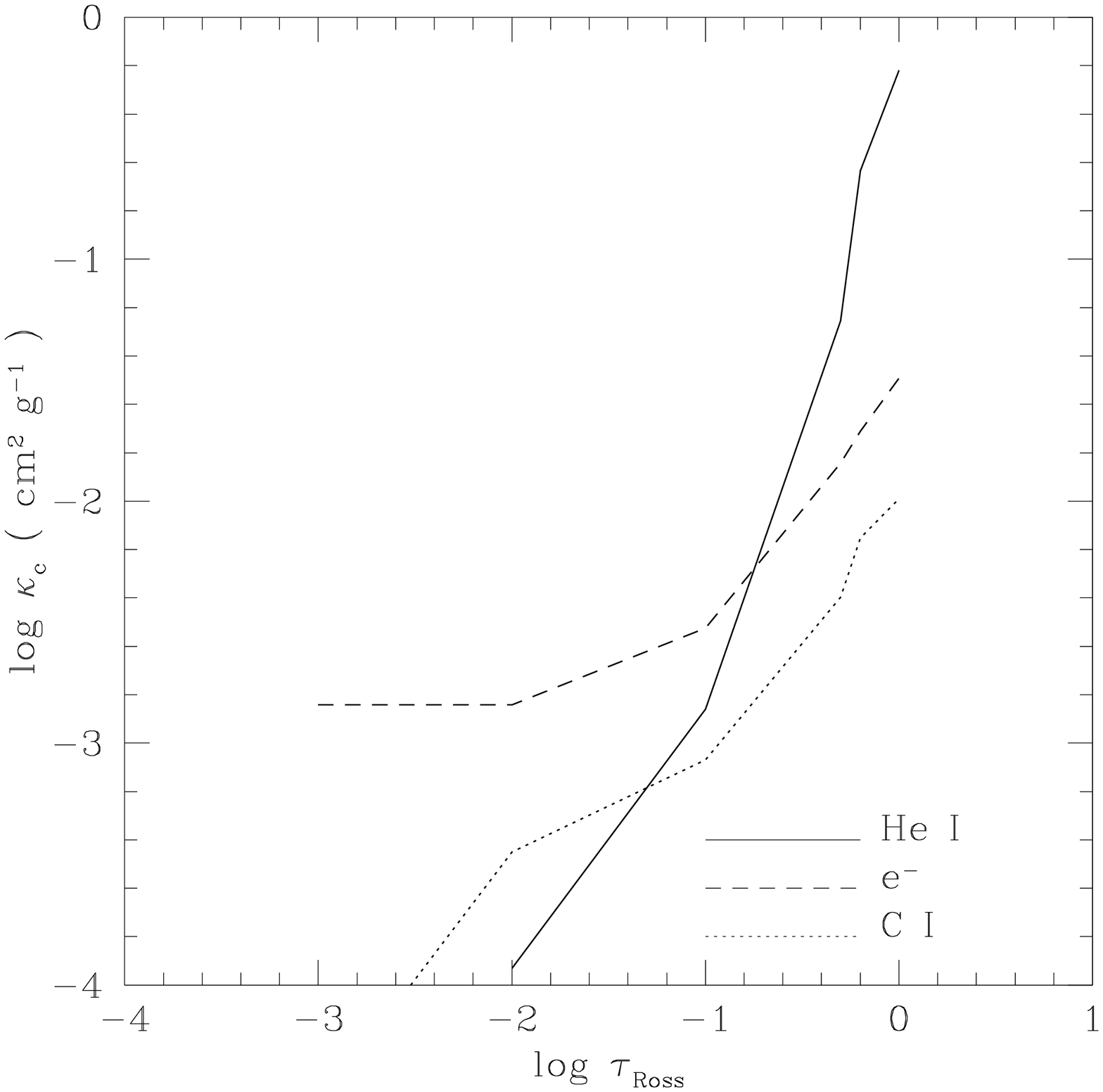}
\caption{Dominant sources of continuous opacity at 5000 \AA~($\kappa_c$) as a function
of Rosseland mean optical depth ($\tau_{Ross}$) for the model atmosphere:
\Teff~= 11000 K, $\log g$ = 1.0 and C/He = 1\%.}
\end{figure}

\begin{figure}
\epsfxsize=8truecm
\epsffile{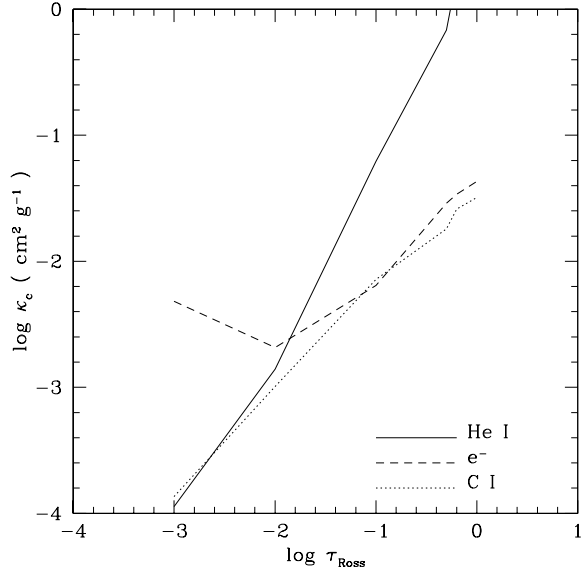}
\caption{Dominant sources of continuous opacity at 5000 \AA~($\kappa_c$) as a function
of Rosseland mean optical depth ($\tau_{Ross}$) for the model atmosphere:
\Teff~= 13000 K, $\log g$ = 2.0 and C/He = 1\%.}
\end{figure}

\noindent
      Figure 2 shows that
electron scattering and photoionization of neutral carbon
 are the major sources of continuous opacity in the line
      forming regions for model of \Teff~= 9500 K.
 Most of the carbon is singly ionized  and contributes about half
 of the total free electrons. The remaining half of the free electrons
come from nitrogen and oxygen.
        At log $\tau$ = $-$0.25, photoionization of carbon and helium
 contribute to the continuous opacity
      equally.
      Helium  dominates the
continuous opacity for  log $\tau$ greater than $-$0.25.
Figures 3 and 4 show that electron scattering and photoionization of neutral helium
are the major sources of continuous opacity in the line forming regions at \Teff~= 11000K
and 13000K.
Photoionization of neutral helium contributes almost all of the total free electrons.
      For hotter stars, e.g., BD --1$^\circ$ 3438 and
      LS IV --1$^\circ$ 002, it is evident
      from Figures 3 and 4 that helium controls the continuum opacity.

Predicted equivalent widths
for representative lines of carbon and helium are illustrated in Figure 5.
This shows several points that were anticipated from Figures 2, 3, and 4.
For cool stars, say \Teff~$\leq 8000$K, the \ion{C}{i} equivalent
 widths are almost independent
of the assumed C/He ratio. The equivalent widths are also almost
independent of effective temperature and surface gravity as a result of the
very similar excitation potentials for the lower levels of the lines and
the photoionization edges. It follows that the predicted
equivalent widths of weak \ion{C}{i} lines are essentially independent of
the atmospheric parameters including  the C/He ratio (Sch\"{o}nberner
1975). This prediction is verified by the observation
that a  \ion{C}{i} line has a similar equivalent width in all \rcb stars
and the coolest of our EHes (see Figure 1 in Rao \& Lambert 1996).
Strong \ion{C}{i} lines are dependent on the 
assumed microtubulence, and, of course, an
equivalent width depends on atomic data, specifically the
line's $gf$-value. 
The \ion{C}{ii} equivalent widths are also almost independent of
effective temperature, surface gravity, and the assumed C/He ratio
for cool stars (see Figure 5).
 A detailed comparison of predicted and observed \ion{C}{i} equivalent
widths of the \rcb stars reveals a systematic discrepancy: the
observed lines are appreciably weaker than predicted. Asplund et al. (2000)
dub this issue `the carbon problem' and review several possible
explanations for it. 
The \ion{He}{i} lines are
dependent on the C/He ratio and very sensitive to the assumed effective
temperature, as the Figure 5 shows. In the case of
stars like \rcb (\Teff~$\simeq 6800$K), the \ion{He}{i} lines are too
weak (and blended) to be used to determine the C/He ratio.

\begin{figure}
\epsfxsize=8truecm
\epsffile{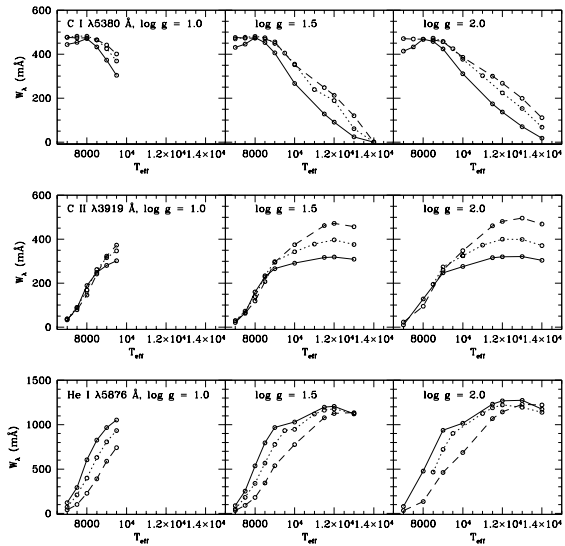}
\caption{Predicted equivalent widths of \ion{C}{i}, \ion{C}{ii}, and \ion{He}{i} lines plotted against $T_{\rm eff}$ for log $g$ values of 1.0, 1.5 and 2.0.
The solid, dotted and dashed lines represent models with C/He of 0.3\%, 1.0\% and 3.0\%, respectively.}
\end{figure}

At the hot limit of our calculations, a different situation pertains.
It is the \ion{He}{i} lines that are independent of the C/He ratio; photoionization
of helium is the dominant contributor to the continuous opacity. The \ion{C}{i}
lines are  weak, and sensitive to effective temperature and surface gravity; 
carbon is predominantly singly-ionized.
The \ion{C}{ii} lines are sensitive to  the C/He ratio and effective temperature
but rather insensitive to the surface gravity. The temperature
sensitivity of \ion{C}{ii} and \ion{He}{i} lines is minimized around 12000K. At high
temperatures, say \Teff~$\geq 12000$K, the \ion{He}{i} equivalent widths
are independent of the assumed C/He ratio and insensitive to the
other atmospheric parameters including the microturbulence; helium
atoms have a large thermal velocity. A check on the models is then
possible by comparing predicted and observed equivalent widths of the
\ion{He}{i} lines. At higher temperatures, helium becomes
ionized and the equivalent width of a \ion{He}{i} line declines as  the
\ion{He}{ii} lines increase in strength.

\section{Abundance analysis}

    The analysis involves the determination of
    \Teff, surface gravity ($\log g$), and microturbulence ($\xi$)
 before estimating the photospheric
 elemental abundances of the star.\\

The microturbulence is derived by requiring that lines of all strengths
      for a particular species
      give the same value of abundance. The derived $\xi$ is found to be independent of
      \Teff, $\log g$ and C/He, adopted for the model atmosphere (Pandey 1999).

      For FQ Aqr, LS IV --14$^\circ$ 109 and BD --1$^\circ$ 3438, we used
      \ion{Fe}{ii}, \ion{Ti}{ii}, \ion{Cr}{ii} and \ion{C}{i} lines, 
and for LS IV --1$^\circ$ 002, we used
      \ion{Fe}{ii}, \ion{S}{ii}, \ion{N}{ii} and \ion{C}{i} lines for determining the
      microturbulent velocity $\xi$.
      In the case of LS IV --14$^\circ$ 109,
      \ion{Fe}{ii}, \ion{Ti}{ii} and \ion{Cr}{ii} lines gave $\xi$ = 6 km s$^{-1}$,
      and \ion{C}{i} lines a value of $\xi$ = 7 km s$^{-1}$.
        We adopt the average microturbulent velocity 
$\xi$ = 6.5 $\pm$ 0.5 km s$^{-1}$ (see Figure 6).
      For FQ Aqr, BD --1$^\circ$ 3438 and LS IV --1$^\circ$ 002, we estimate the
      microturbulent velocity $\xi$ as 7.5 $\pm$ 0.5 km s$^{-1}$,
      10.0 $\pm$ 1.0 km s$^{-1}$ and 10.0 $\pm$ 1.0 km s$^{-1}$, respectively.
The microturbulence values provided by the different elements agree
well within the errors quoted above.\\

\begin{figure}
\epsfxsize=8truecm
\epsffile{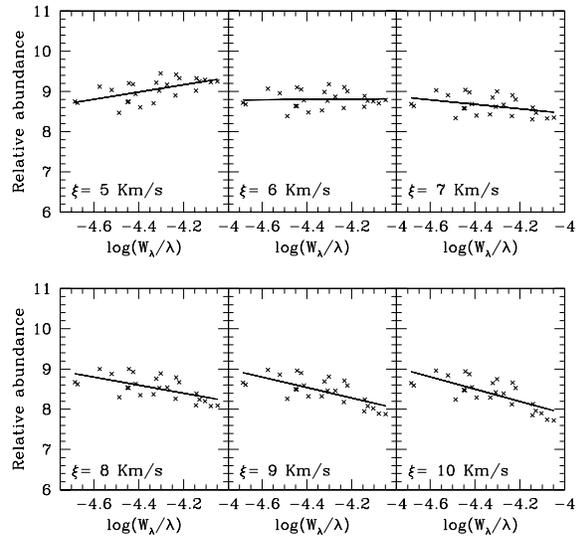}
\caption{Relative abundances from \ion{Ti}{ii} lines for LS IV --14$^\circ$ 109, plotted against their line strength,
represented by (log W$_{\lambda}$/$\lambda$) for different values of $\xi$.}
\end{figure}
%
%
%
  \Teff~is estimated by requiring that the
      lines of a particular species but of differing excitation potentials
        should return the same elemental abundance.
The model grid is searched for the model that satisfies this
condition. The optimum \Teff~is found to be independent of 
the adopted $\log g$ and C/He for the model atmosphere (Pandey 1999).

In all cases, \ion{Fe}{ii} lines, which are numerous and span a range of
excitation potentials, were used to determine \Teff. Figure 7
illustrates the procedure used to determine \Teff.
No other species shows such a large range in excitation potential
to determine \Teff.\\

\begin{figure}
\epsfxsize=8truecm
\epsffile{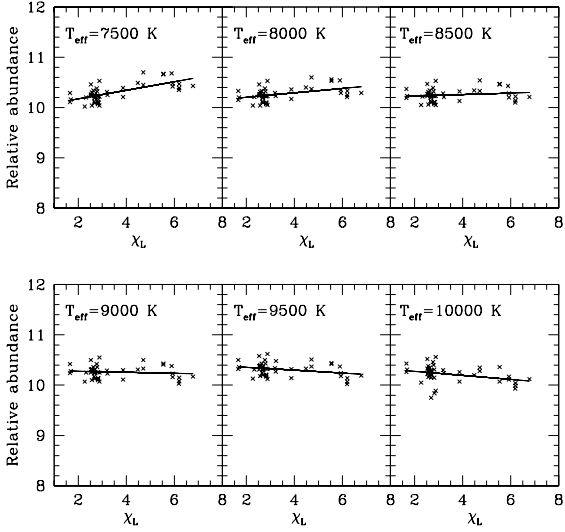}
\caption{Excitation balance for FQ Aqr using \ion{Fe}{ii} lines.}
\end{figure}

  The surface gravity is estimated by the requirement that
 the model atmosphere
      gives the same  abundances for
      neutral, singly, and doubly ionized lines of a given element.
 The ionization
     equilibria are independent of the adopted C/He of the model atmosphere.\\

      Since the ionization equilibrium depends on both
      surface gravity and temperature, its imposition defines a locus in the
      $\log g$ -- \Teff~plane
for a given pair of ions (or atom and ion) of an element.
            The ionization equilibrium of the following species
      (when sufficient lines are available) are used to estimate
      \Teff and $\log g$:
      \ion{S}{ii}/\ion{S}{i}, \ion{Si}{iii}/\ion{Si}{ii}/\ion{Si}{i}, \ion{N}{ii}/\ion{N}{i},
      \ion{Al}{iii}/\ion{Al}{ii}/\ion{Al}{i}, \ion{C}{ii}/\ion{C}{i}, 
      \ion{Fe}{iii}/\ion{Fe}{ii}/\ion{Fe}{i}, \ion{Mg}{ii}/\ion{Mg}{i} 
      and \ion{O}{ii}/\ion{O}{i}.
The solutions for \Teff~and $\log g$ obtained for
LS IV --14$^\circ$ 109 are shown in
Figure 8, where the ionization equilibria are shown by different
line-types and the excitation balance by arrow heads in the \Teff~-- $\log g$ plane.
%

\begin{figure}
\epsfxsize=8truecm
\epsffile{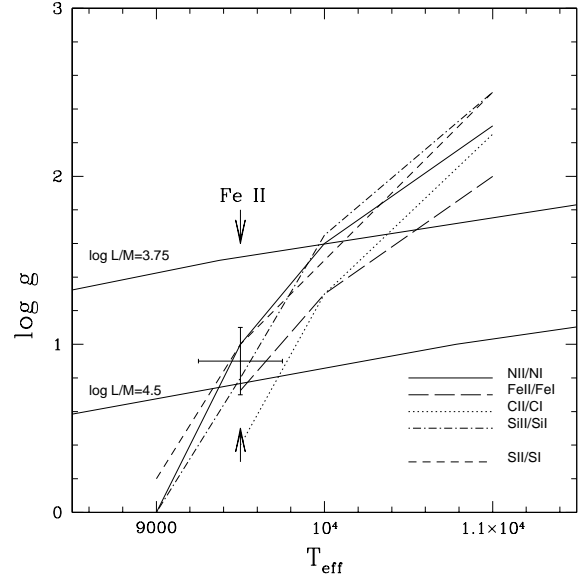}
\caption{Final \Teff~and $\log g$ of LS IV --14$^\circ$ 109 with error bars.}
\end{figure}

%
%
If there is an evolutionary link between our cool EHes and the hot EHes and the
\rcb stars, the mass $M$ and luminosity $L$ of the stars in the
 three groups should
be related. One might suppose that for a particular isochrone,
 the mass is constant.
In essence, one assumes a close relation between $M$ and $L$; it is readily
shown that the ratio $M/L \propto g/T_{eff}^4$. Hence, as an additional
locus, we show
 lines of constant log$(L/M)$ on Figure 8; the chosen values of
 3.75 and 4.5 span the values of a majority of
the hot EHe stars (Jeffery 1996) but a few stars have
a lower log$(L/M)$. 
 Note that we have not used these lines of constant log$(L/M)$
in determining the \Teff~and $\log g$.

Since available loci from ionization equilibrium run
approximately parallel to each other, separate solutions
for \Teff~and log $g$ cannot be obtained.
 Excitation equilibrium provides a line in the
\Teff~vs log $g$ plane that intersects the ionization
equilibria at a steep angle enabling  log $g$ to be
estimated. The adopted \Teff~and log $g$ is indicated on
Figure 8.

      The $gf$-values and excitation potentials for the lines used in our LTE analysis were
taken from the compilations by R. E. Luck (private communication), Jeffery (1994),
Th\'{e}venin (1989, 1990) and Kurucz \& Peytreman (1975).
      We have used the $gf$-values of \ion{C}{i} lines from Opacity project (Seaton et al. 1994; Luo \& Pradhan 1989;
      Hibbert et al. 1993).
      The Stark broadening
      and radiation broadening coefficients were mostly taken from the
      compilation by Jeffery (1994).
      The data for computing \ion{He}{i} profiles are obtained from various sources.
     The $gf$-values are taken from
      Jeffery (1994),
      radiation broadening coefficients from Wiese et al. (1966),
      and electron
      broadening coefficients from the combination of
      Griem et al. (1962), Bassalo, Cattani, \& Walder (1980),
       Dimitrijevic \& Sahal-Brechot (1984), and Kelleher (1981).
        The effects of ion broadening are also included.

         All the synthesized spectra corrected for the instrumental profile were convolved with a Gaussian profile to
      give a good fit to typical unblended line profiles. Only weak and unblended lines of trace elements were
      used to fix the full width at half maximum (FWHM) of the Gaussian profile. We find that the
      resultant FWHM of the Gaussian profile of the stellar lines used is more than
      the FWHM of the instrumental profile. We attribute
      this extra broadening to a combination of rotation and macroturbulence (see $v_{\rm M}$ in Table 1).

\begin{table*}
\caption{Final stellar parameters for cool extreme helium stars}
\begin{center}
\begin{tabular}{lcccccc} \hline
 Star                  &       \Teff    &        $\log g$    &        $\xi$&C/He&R.V.& $v_{\rm M}$\\
                       &       K    &   cgs units     &    km s$^{-1}$&\%&km s$^{-1}$& km s$^{-1}$\\ \hline
 FQ Aqr                & 8750$\pm$250  & 0.75$\pm$0.25 & 7.5$\pm$0.5&0.5&16$\pm$3 (59)& 20            \\
 LS IV --14$^\circ$ 109  & 9500$\pm$250  & 0.90$\pm$0.20 & 6.5$\pm$0.5&1.0&5$\pm$2 (47)& 15            \\
 BD --1$^\circ$ 3438     & 11750$\pm$250  & 2.30$\pm$0.40 & 10$\pm$1.0&0.2&$-$22$\pm$2 (45)& 15           \\
 LS IV --1$^\circ$ 002   & 12750$\pm$250  & 1.75$\pm$0.25 & 10$\pm$1.0&0.6&$-$20$\pm$3 (22)& 20           \\ \hline
\end{tabular}
\end{center}
\end{table*}
 
The stars FQ~Aqr and LS IV --14$^\circ$ 109 were earlier analysed by 
 Asplund et al. (2000) based on 30,000 resolution spectra obtained at CTIO
 in the wavelength region 5500\AA\ to 6800\AA. The present analysis is based on new
 higher resolution spectra covering 3800\AA\ to 10000\AA\ providing more 
 spectral lines of many more species.

\subsection{FQ Aqr}

The \ion{Fe}{ii} lines require that
\Teff~= 8750 $\pm$ 250K. Lines stronger than 200 m\AA\ were rejected.
Ionization equilibria involving N, Al, Mg, S, and Fe were
considered.
The ionization balance of \ion{S}{ii}/\ion{S}{i} is given highest
weight because the lines of \ion{S}{ii} and \ion{S}{i} identified in the spectra
are weak. \ion{C}{ii}/\ion{C}{i} ionization balance is given a lower
weight because of the potential carbon problem (see below).
A lower weight is also given to the
ionization balance of \ion{Al}{ii}/\ion{Al}{i} because we have only one line of \ion{Al}{i}
and the $gf$-values for \ion{Al}{ii} lines may be unreliable.
Equal weights are given to the ionization balance of 
\ion{N}{ii}/\ion{N}{i}, \ion{Mg}{ii}/\ion{Mg}{i} and \ion{Fe}{ii}/\ion{Fe}{i}.
The spread in these loci is similar to that found for the hot EHes - see, for example, Jeffery (1998).
We adopt \Teff~= 8750 $\pm$ 250K and log $g = 0.75 \pm0.25$ (Table 1).

At the \Teff~of FQ Aqr, carbon is predicted to be the leading contributor
to the continuous opacity (Figure 2). Under these circumstances, the \ion{He}{i}
lines are sensitive to the C/He ratio (Figure 5) which may be derived by fitting
the \ion{He}{i} lines at 5048, 5876, and 6678\AA\ (Figure 9).
The lines give C/He = 1.2$\pm$0.2\% where the uncertainty reflects only the scatter
of the three results. Temperature and gravity errors contribute about
$\pm$0.7\% to the C/He ratio. The error in microturbulence contributes a negligible 
amount of uncertainty to the C/He ratio.

For the adopted model, the predicted 
equivalent widths of \ion{C}{i} and \ion{C}{ii} lines are almost independent of the model's 
assumed C/He ratio as long as C/He exceeds the minimum
necessary for carbon to dominate the continuous opacity (C/He $\geq$ 0.5\%).
Then there are essentially no free parameters with which to adjust the
predicted equivalent width of a carbon line with a given $gf$-value. Comparison of
predicted and observed equivalent widths of FQ Aqr shows, as it did for the 
\rcb stars (Asplund et al. 2000), that observed equivalent widths of the \ion{C}{i}
lines are weaker than predicted. For example, models with the adopted \Teff~and
log $g$ require the C abundance (or $gf$-value) to be reduced by about 0.4 dex for
both C/He = 3\% and 1\% models.
However for the C/He = 0.3\% model,
the difference is a mere 0.06 dex. At this
C/He ratio, carbon is not the dominant source of continuous opacity, instead 
electron scattering (most of the free electrons are coming from nitrogen and oxygen) 
is the major source of continuum opacity.
To match the observed equivalent widths of the \ion{C}{ii} lines, models with the adopted \Teff~and log $g$ 
require the C abundance (or $gf$-value) to be reduced by about 0.2 dex for 
both C/He = 3\% and 1\% models.
At C/He = 0.3\%, predicted and observed
equivalent widths agree to better than 0.05 dex. Temperature and gravity errors contribute
about $\pm$0.3 dex to the C/He ratio derived using \ion{C}{ii} lines.

If C/He $\geq$ 0.3\%, there is a carbon problem whose resolution is presumably closely related to the unidentified
solution to the carbon problem of the \rcb stars. The problem is larger for \ion{C}{i} than the \ion{C}{ii} lines,
a result also found for the \rcb stars.
The magnitude of the problem at a given C/He ratio
 is smaller than for the \rcb stars and vanishes at a higher C/He ratio
than for the \rcb stars. To within the uncertainties allowed by the
model atmosphere parameters, a ratio
C/He $\approx$ 0.5\% 
produces a tolerable fit to the \ion{He}{i}, \ion{C}{i}, and \ion{C}{ii} lines.
The final abundances as given in Table 2 are derived for C/He = 0.5\%.

%
%

\begin{figure}
\epsfxsize=8truecm
\epsffile{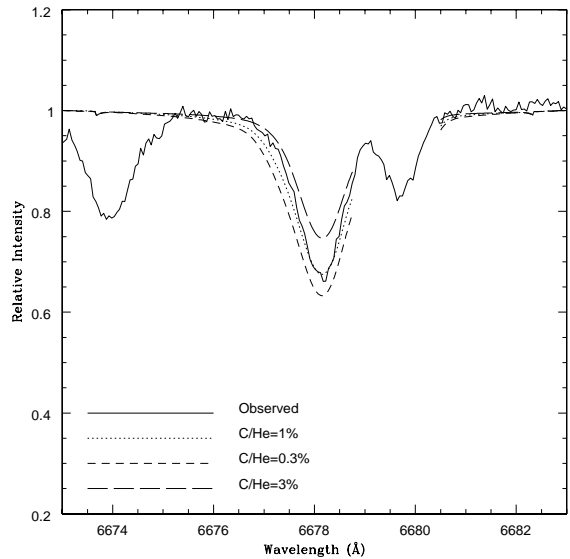}
\caption{Observed and synthesized \ion{He}{i} $\lambda$ 6678 \AA~line profile of FQ Aqr. The \ion{He}{i}
line profiles are synthesized
using models of \Teff~= 8750 K and log $g$ = 0.75 but for different values of C/He.}
\end{figure}

\subsection{LS IV --14$^\circ$ 109}

Adopted atmospheric parameters (Table 1) are based on the excitation
and ionization equilibria shown in Figure 8.
The \Teff~= 9500$\pm$250K is provided from the
excitation of \ion{Fe}{ii} lines.
Loci corresponding to ionization equilibria for five sets of atoms
and ions  are remarkably consistent; three are essentially
identical. 
 Equal weights
are given 
to the ionization balance of \ion{N}{ii}/\ion{N}{i}, \ion{Fe}{ii}/\ion{Fe}{i}, \ion{Si}{ii}/\ion{Si}{i}
 and \ion{S}{ii}/\ion{S}{i}.

The C/He ratio is found from a fit to \ion{He}{i} profiles. We have synthesised
the 3872\AA, and 5048\AA\ lines. Unfortunately, the 5876\AA, and 6678\AA\
are not on our spectra, and the 7065\AA\ line is blended with telluric
H$_2$O lines. A value C/He = 1.2$\pm$0.2\% is obtained.
Temperature and gravity errors contribute about
$\pm$0.6\% to the C/He ratio. The error in microturbulence contributes a negligible 
amount of uncertainty to the C/He ratio.
 
%
The \ion{C}{i} lines as analysed with the C/He = 1\% model give an abundance
 corresponding to C/He = 0.75$\pm$0.3\%.
When uncertainties attributable to the atmospheric parameters (temperature, gravity and microturbulence) are
considered, this result  is consistent
with that derived from the \ion{He}{i} lines.
Analysis of the \ion{C}{ii} lines with C/He = 1\% model gives
C/He = 0.9$\pm$0.25\% for the adopted model where the errors
reflect the uncertainty in the estimated \Teff~and log $g$.
The abundances are derived for C/He = 1\% which provides an acceptable fit to the \ion{He}{i},
\ion{C}{i} and \ion{C}{ii} lines.

\subsection{BD --1$^\circ$ 3438}

In deriving the atmospheric parameters (Table 1)
equal weights are given
      to the ionization balance of \ion{N}{ii}/\ion{N}{i}, \ion{Fe}{ii}/\ion{Fe}{i} and
      \ion{C}{ii}/\ion{C}{i}.
Helium is the
major contributor of continuum opacity and, hence, the \ion{He}{i} equivalent
widths are almost independent of the C/He ratio.
The observed profiles of
the \ion{He}{i} 5048\AA\ and 5876\AA\ lines are well
fit by the predictions for the \Teff~= 11500K and log $g$ = 2.0
model within errors (Figure 10); there is no helium problem
analogous to the carbon problem as seen for \rcb stars.

\begin{figure}
\epsfxsize=8truecm
\epsffile{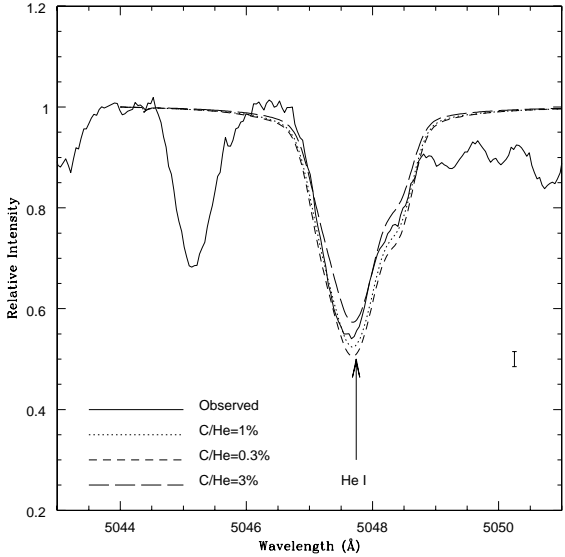}
\caption{Observed and synthesized \ion{He}{i} $\lambda$ 5047.74 \AA~line profile of BD --1$^\circ$ 3438. The \ion{He}{i} line profiles ar
e synthesized
using model of \Teff~= 11500 K and log $g$ = 2.0, for different values of C/He. The uncertainty on the ordinate is shown by the error bar.}
\end{figure}

%
With He providing the continuous opacity, the \ion{C}{i} and \ion{C}{ii} lines
provide a  measure of  the C/He ratio. The \ion{C}{i} lines give
log $\epsilon(\rm C)$ = 8.95$\pm$0.27.
Two \ion{C}{ii} lines give 8.8$\pm$0.08.
 The \ion{C}{i} and \ion{C}{ii} lines
are in fair agreement and imply a C/He = 0.2$\pm$0.03\%. Temperature and gravity errors are included. By definition, the carbon problem is not apparent for 
atmospheres for which carbon is not the dominant opacity source.

\subsection{LS IV --1$^\circ$ 002}

 The adopted \Teff~and $\log g$ using the excitation
       and ionization balance criteria  are given in
       Table 1.
The derived \Teff~from the excitation balance of \ion{Fe}{ii} lines
 is 12750$\pm$250K.
Ionization equilibria for six elements are available.
 Equal weights are given to the ionization balance of \ion{O}{ii}/\ion{O}{i}, \ion{N}{ii}/\ion{N}{i},
 \ion{Fe}{iii}/\ion{Fe}{ii} and \ion{C}{ii}/\ion{C}{i}. Smaller weights are given to the ionization
 balance of \ion{Si}{iii}/\ion{Si}{ii} and \ion{Al}{iii}/\ion{Al}{ii} because the lines available
 are very few and the sources of $gf$-values for these lines are not
reliable.

The \ion{He}{i} profiles are effectively independent of the C/He ratio -
see Figure 11
 where we show good agreement between the predicted and observed 5048\AA\ profiles.

\begin{figure}
\epsfxsize=8truecm
\epsffile{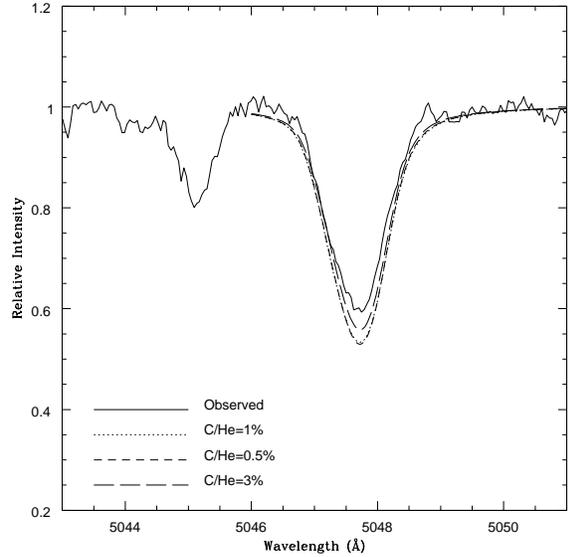}
\caption{Observed and synthesized \ion{He}{i} $\lambda$ 5047.74 \AA~line profile of LS IV --1$^\circ$ 002. The \ion{He}{i} line profiles ar
e synthesized
using model of \Teff~= 13000 K and log $g$ = 2.0, for different values of C/He.}
\end{figure}

The \ion{C}{i} lines give a C abundance equivalent to C/He = 0.6$\pm$0.2\%.
 At this temperature, the spectrum of \ion{C}{ii} is well
represented: 7 lines give C/He = 0.6$\pm$0.2\%. Temperature and gravity errors are
included. We adopt C/He = 0.6\%.

\section{Abundances}

      The final abundances were calculated using models of C/He = 0.5\%, 1.0\%, 0.2\%, and 
0.6\% for FQ Aqr, LS IV --14$^\circ$ 109, BD --1$^\circ$ 3438 and LS IV --1$^\circ$ 002, respectively.
Detailed line lists are obtainable by anonymous ftp to astro.as.utexas.edu from pub/pandey/linelist.ps.

   For the adopted stellar parameters,
the individual elemental abundances listed in Table 2 are given as log $\epsilon(i)$, normalized
such that log $\Sigma$$\mu_i \epsilon(i)$ = 12.15. The abundances determined from the neutral 
and ionized species of an element are separately listed. The number of lines used in our analysis is given
within brackets.
The errors quoted are mainly due to line-to-line scatter of the abundances.
   The error in abundances due to the uncertainty in the adopted stellar parameters is discussed in Appendix A.
Solar abundances from Grevesse, Noels, \& Sauval (1996) are also shown in Table 2.  

Hydrogen is present in the atmospheres of these stars.
 We have identified four Balmer lines (H$\alpha$, H$\beta$,
H$\gamma$, and H$\delta$) in our spectra.
 Our H abundance is based on the  H$\alpha$ line but H$\beta$ gives a consistent
result. We have not used the  H$\gamma$ and H$\delta$ lines  because they are
blended and noisy. Line broadening for Balmer lines was
considered in determining the H abundance.

The difference in abundances reported by Asplund et al. (2000) and our analyses of FQ~Aqr and LS IV --14$^\circ$ 109
is essentially due to differences in adopted stellar parameters. 

\begin{table*} 
\caption{The individual elemental abundances derived for each ion for the analyzed EHe stars} 
\begin{center} 
\begin{tabular}{llllll}
\hline 
 &FQ Aqr&LS IV --14$^\circ$ 109&BD --1$^\circ$ 3438&LS IV --1$^\circ$ 002&Sun\\ 
\hline \Teff~(K) &8750  &9500  &11750  &12750&  \\ 
$\log g$ (cgs units) &0.75 &0.9 &2.3&1.75& \\
 C/He              &0.5\% &1\% &0.2\%&0.6\% & \\ 
$\xi$ (km s$^{-1})$&8.0 &6.0 &10.0 &10.0& \\ 
\hline H~I&6.2(1)&6.2(1)&5.6(1)&7.1(1)&12.0\\ 
He~I&11.54(3)&11.54(2)&11.54(2)&11.54(1)&10.99\\ 
C~I &9.0$\pm$0.14(30) &9.4$\pm$0.16(25)&9.0$\pm$0.17(28)&9.3$\pm$0.15(15)&8.55\\ 
C~II & 9.0$\pm$0.10(2) & 9.5$\pm$0.06(3)& 8.8$\pm$0.06(2)& 9.3$\pm$0.13(7)&    \\ 
N~I& 7.1$\pm$0.20(5)& 8.6$\pm$0.28(10)& 8.4$\pm$0.20(11)& 8.2$\pm$0.10(6)&7.97\\ 
N~II& 7.2$\pm$0.07(2)& 8.6$\pm$0.18(5)& 8.6$\pm$0.15(9)& 8.3$\pm$0.20(14)&\\ 
O~I& 8.9$\pm$0.15(8) & 8.5$\pm$0.16(8)& 8.4$\pm$0.19(6)& 8.8$\pm$0.15(3)&8.87\\ 
O~II& ... & ...& ...& 8.9$\pm$0.05(5)&\\ 
Ne~I& 7.9$\pm$0.25(14)&        9.4$\pm$0.26(13)&       8.8$\pm$0.14(9)&       9.0$\pm$0.13(9)&8.10\\ 
Na~I& 5.5$\pm$0.26(2)&        6.8$\pm$0.22(4)&       6.3(1) &       6.5(1) &6.33\\
Mg~I& 5.5$\pm$0.12(5)&6.9$\pm$0.26(2)& ...& ...&7.58\\
Mg~II& 6.0$\pm$0.11(6)&7.3$\pm$0.10(3)&6.9$\pm$0.03(3)& 6.9$\pm$0.23(6)&\\
Al~II& 4.7$\pm$0.18(4)& 7.1$\pm$0.2(7) & 6.0$\pm$0.20(8) & 5.4$\pm$0.17(8)&6.47\\
Al~III& ...& 6.7$\pm$0.10(2)& ... & ... &\\
Si~I& ...& 7.6$\pm$0.21(5)& ...& ...&7.55\\
Si~II& 6.3$\pm$0.24(6) & 7.8$\pm$0.24(3)& 6.5$\pm$0.14(5) & 5.9$\pm$0.10(3) &\\
P~II& 4.2$\pm$0.23(2)& 5.3$\pm$0.27(3)&       5.3$\pm$0.23(3)&      5.1$\pm$0.11(3)&5.50\\
S~I& 6.1$\pm$0.15(3)& 7.6$\pm$0.24(4)& ...& ...&7.23\\
S~II& 5.9$\pm$0.16(7)& 7.5$\pm$0.40(12)& 6.9$\pm$0.20(18)& 6.7$\pm$0.20(35)&\\
Ca~I& 4.0(1) & 5.5(1) & ...& ...&6.36\\
Ca~II& 4.2(1) & 5.6$\pm$0.22(2) & 5.5$\pm$0.10(2) & 5.8$\pm$0.05(2) &\\
Sc~II&      2.1$\pm$0.12(7)&        3.3$\pm$0.21(5)& ...       & ...     &3.17  \\
Ti~II&       3.2$\pm$0.25(42)&        4.3$\pm$0.21(25)&       4.6$\pm$0.28(18)&       4.7$\pm$0.14(5)&5.02\\
Cr~I& 4.0(1)& ...& ...& ...       &5.69\\
Cr~II& 3.6$\pm$0.13(30)& 5.1$\pm$0.24(30)& 4.9$\pm$0.27(23)&  ...      &\\
Mn~II&      4.3$\pm$0.23(3)&        5.3$\pm$0.27(11)&       5.1$\pm$0.10(4)& ...       &5.47\\
Fe~I& 5.1$\pm$0.12(7)& 6.8$\pm$0.17(7)& 7.1$\pm$0.04(4)& ...&7.50\\
Fe~II& 5.4$\pm$0.12(59)& 7.0$\pm$0.21(47)& 6.7$\pm$0.20(45)& 6.3$\pm$0.12(22)&\\
Fe~III& ...& ...& ...& 6.1$\pm$0.22(2)&\\
Ni~I& ...   &       6.6$\pm$0.17(4)&   ...     &  ...     &6.25 \\
Sr~II&       0.5$\pm$0.03(2)&       2.6$\pm$0.05(2)&       2.8$\pm$0.03(2) &       2.7$\pm$0.20(2) &2.97\\
Y~II&  ...     &       1.9$\pm$0.17(3)&  ...      &   ...     &2.24\\
Zr~II&      0.8$\pm$0.24(2)&       1.9$\pm$0.10(2)& ...       &  ...      &2.60\\
Ba~II&      0.5(1)&       1.7$\pm$0.13(3)&   ...       &   ...       &2.13\\ \hline
\end{tabular}
\end{center}
\end{table*}

\section{Cool EHe, Hot EHe, and R CrB Stars -- Compositions}

The chemical compositions of the cool EHes are clues to the history
of these enigmatic stars, which assuredly began life as normal stars with a
H-rich atmosphere. Our discussion begins with a comparison of the
cool EHes with their putative relatives the hot EHes and the R CrB
stars.
Data on the composition of hot EHes are taken from Jeffery's (1996)
review and subsequent papers; Harrison \& Jeffery (1997), Jeffery \& Harrison (1997), Drilling, Jeffery \&
Heber (1998),
Jeffery (1998), Jeffery et al. (1998) and Jeffery, Hill, \& Heber (1999). We exclude from consideration the
hydrogen-deficient binaries such as $\upsilon$\,Sgr, which have normal carbon abundances.
Chemical compositions of the \rcb stars are taken from Asplund et al. (2000).
Our goal is to elucidate similarities, differences, and trends.
It has, of course, to be kept in
mind that not only evolutionary associations but also systematic errors
may provide variable signatures across the collection of stars running
from extreme helium stars as hot as \Teff~= 30,000K to cool
\rcb stars with \Teff = 6000K, and with surface gravities from
log $g \simeq$ 1 to 4.

Mean abundances for the three groups are given in Table 3 where the
dispersion $\sigma$ is a measure of the range of the published
abundances. For the \rcb stars, the given carbon abundance is the spectroscopic
carbon abundance which is 0.6 dex lower than the input carbon abundance used
for the model. We give both log$\epsilon (el)$ and [$el$/Fe]. From this table,
we exclude FQ Aqr, four minority class R~CrBs, and the hot \rcb DY~Cen (Jeffery \& Heber 1993) because these stars show
a much lower Fe abundance and, in some cases, other striking abundance
anomalies. The minority class R~CrBs show lower Fe abundance and higher Si/Fe and S/Fe
ratios than majority class R~CrBs (Asplund et al. 2000). Abundances for these stars are summarized in Table 4. 
Models of C/He = 1\% are used to
derive the abundances of \rcb stars except for V854~Cen for which C/He = 10\% model is used (see Asplund et al. 1998).
 
The mean abundances of $\alpha$-elements relative to Fe, for the three groups are not as expected for
their mean Fe abundances (Table 3).
Relative to sulphur, mean abundances for the three groups are given in Table 5 for
elements from sodium through to the iron group which were measured in 40 per cent or more of the stars comprising
each group. Sulphur is chosen as the reference element in preference
to iron, the customary choice, for reasons outlined below. The
dispersion $\sigma$ given in parentheses for each entry is likely
dominated by the measurement errors.
Table 5 also gives the solar ratios and those expected of a normal
star with [Fe/H] = --1 (Wheeler, Sneden, \& Truran 1989; Lambert 1989;
Goswami \& Prantzos 2000). We have assumed that the abundances of the $\alpha$-elements
Mg, Si, S, Ca and Ti vary in concert with decreasing [Fe/H]. There are no observations
of P in stars but we assume [P/Fe] increases less steeply than [S/Fe] with decreasing
[Fe/H]. FQ Aqr is included under cool EHes in Table 5.

\begin{table*}
\begin{center}
\caption{The mean abundances log $\epsilon$($el$) and the mean abundance ratios
[$el$/Fe] for cool EHe (excluding FQ Aqr), hot EHe (excluding V652 Her and
HD 144941) and majority class \rcb stars. The dispersion $\sigma$
and the number of stars (\#) are also given.}
\begin{tabular}{lrrccrrccrrcc} \hline
\multicolumn{1}{c}{Element($el$)} & \multicolumn{4}{c}{cool EHe stars}
& \multicolumn{4}{c}{hot EHe stars} & \multicolumn{4}{c}{majority \rcb stars}\\ \cline{2-4}
\cline{6-8} \cline{10-12}
& \multicolumn{1}{c}{log $\epsilon$($el$) ($\sigma$)}
& \multicolumn{1}{c}{[$el$/Fe] ($\sigma$)}
& \multicolumn{1}{c}{\#} &
& \multicolumn{1}{c}{log $\epsilon$($el$) ($\sigma$)}
& \multicolumn{1}{c}{[$el$/Fe] ($\sigma$)}
& \multicolumn{1}{c}{\#} &
& \multicolumn{1}{c}{log $\epsilon$($el$) ($\sigma$)}
& \multicolumn{1}{c}{[$el$/Fe] ($\sigma$)}
& \multicolumn{1}{c}{\#} &\\ \hline
H & 6.3 (0.8) & $-$4.8 (1.2) & 3 && 8.0 (0.5)  & $-$3.4 (0.7) & 10 && 6.1 (1.0) & $-$4.9 (1.1) & 13&\\
C & 9.2 (0.4) & 1.6 (0.6)    & 3 && 9.3 (0.2)  &    1.5 (0.3) & 10 && 8.9 (0.2) &    1.4 (0.2) & 14&\\
N & 8.5 (0.2) & 1.4 (0.3)    & 3 && 8.3 (0.4)  &    0.9 (0.3) & 10 && 8.6 (0.2)  &    1.6 (0.3) & 14&\\
O & 8.6 (0.3)  & 0.6 (0.7)    & 3 && 8.6 (0.3)  &    0.3 (0.6) & 10 && 8.2 (0.5)  &    0.3 (0.6) & 14&\\
Ne& 9.1 (0.3)  & 1.8 (0.5)    & 3 && 9.2 (0.3)  &    1.5 (0.3) & 3  && 8.3       &    1.2      & 1 &\\
Na& 6.5 (0.3)  & 1.1 (0.5)    & 3 &&  ...     &  ...         &... && 6.1 (0.2)  &    0.8 (0.1) & 13&\\
Mg& 7.0 (0.2)  & 0.3 (0.4)    & 3 && 7.6 (0.3)  &    0.6 (0.4) & 10 && 6.7 (0.3)  &    0.0 (0.4) & 3 &\\
Al& 6.2 (0.9)  & 0.5 (0.5)    & 3 && 6.1 (0.5)  &    0.2 (0.4) & 10 && 6.0 (0.3)  &    0.5 (0.3) & 12&\\
Si& 7.4 (0.3) & 0.8 (0.4)    & 3 && 7.4 (0.4) &    0.5 (0.4) & 10 && 7.1 (0.2) &    0.6 (0.2) & 14&\\
P & 5.2 (0.1) & 0.6 (0.4)    & 3 && 5.7 (0.5) &    0.8 (0.5) & 10 && 5.9      &    1.4      & 1 &\\
S & 7.0 (0.4) & 0.7 (0.3)    & 3 && 7.1 (0.3) &    0.5 (0.5) & 10 && 6.9 (0.4) &    0.7 (0.3) & 14&\\
Ca& 5.6 (0.2) & 0.1 (0.6)    & 3 && 6.4 (0.4) &    0.5 (0.1) & 7  && 5.4 (0.2) & $-$0.1 (0.2) & 14&\\
Sc& 3.3      & 0.6         & 1 && 3.2 (1.6) &    0.4 (0.9) & 2  && 2.9 (0.3) &    0.7 (0.4) & 4 &\\
Ti& 4.5 (0.2) & 0.4 (0.7)    & 3 && 5.0 (0.8) &    0.3 (0.3) & 3  && 4.0 (0.2) &    0.0 (0.3) & 6 &\\
Cr& 5.0 (0.1) & $-$0.1 (0.0) & 2 && 5.6 (0.8) &    0.4 (0.1) & 2  &&  ...     &   ...       &...&\\
Mn& 5.2 (0.1) & 0.3 (0.0)    & 2 && 5.3 (1.3) &    0.3 (0.6) & 2  &&  ...     &   ...       &...&\\
Fe& 6.6 (0.5) & 0.0         & 3 && 6.9 (0.3) &    0.0      & 10 && 6.5 (0.3) &    0.0      & 14&\\
Ni&  ...     & ...         &...&& 5.8 (1.0) &    0.1 (0.3) & 2  && 5.8 (0.2) &    0.6 (0.2) & 14&\\
Zn&  ...     &  ...        &...&&  ...     &   ...       &... && 4.4(0.3) &    0.7(0.2) & 11&\\
Sr& 2.7 (0.1) & 0.6 (0.5)    & 3 &&  ...     &  ...        & ...&& ...      &   ...       &...&\\
Y & 1.9      & 0.2         & 1 &&  ...     &  ...        & ...&& 2.1 (0.5) &    0.9 (0.4) & 14&\\
Zr& 1.9      & $-$0.2      & 1 &&  ...     &  ...        & ...&& 2.1 (0.4) &    0.4 (0.5) & 6 &\\
Ba& 1.7      & 0.1         & 1 &&  ...     &   ...       & ...&& 1.5 (0.5) &    0.4 (0.4) & 14&\\
\hline
\end{tabular}
\end{center}
\end{table*}

\begin{table*}
\centering
\begin{minipage}{170mm}
\caption{The abundances log $\epsilon$($el$) and the abundance ratios
[$el$/Fe] for minority class \rcb stars (V~CrA, VZ~Sgr, V3795~Sgr, and V854~Cen),
hot \rcb DY~Cen, and cool EHe star FQ Aqr.}
\begin{tabular}{lcrccrcrrccrccrccrc} \hline
\multicolumn{1}{c}{$el$} & \multicolumn{3}{c}{V~CrA}
& \multicolumn{3}{c}{VZ~Sgr} & \multicolumn{3}{c}{V3795~Sgr}
& \multicolumn{3}{c}{V854~Cen} & \multicolumn{3}{c}{DY~Cen} & \multicolumn{3}{c}{FQ Aqr}\\ \cline{2-3}
\cline{5-6} \cline{8-9} \cline{11-12} \cline{14-15} \cline{17-18}
& \multicolumn{1}{c}{abun$^a$} & \multicolumn{1}{c}{[$el$/Fe]} &
& \multicolumn{1}{c}{abun$^a$} & \multicolumn{1}{c}{[$el$/Fe]} &
& \multicolumn{1}{c}{abun$^a$} & \multicolumn{1}{c}{[$el$/Fe]} &
& \multicolumn{1}{c}{abun$^a$} & \multicolumn{1}{c}{[$el$/Fe]} &
& \multicolumn{1}{c}{abun$^a$} & \multicolumn{1}{c}{[$el$/Fe]} &
& \multicolumn{1}{c}{abun$^a$} & \multicolumn{1}{c}{[$el$/Fe]} &\\
\hline
H & 8.0 &$-$2.0&& 6.2 &$-$4.1&& $<$4.1 &$-$6.0&& 9.9 &$-$0.6&&10.8 & 1.3 && 6.2 &$-$3.7& \\
C & 8.6 & 2.1  && 8.8 & 2.0  && 8.8    & 2.2  && 9.6 & 2.6  && 9.5 & 3.5 && 9.0 & 3.0 & \\
N & 8.6 & 2.6  && 7.6 & 1.3  && 8.0    & 1.9  && 7.8 & 1.3  && 8.0 & 2.5 && 7.2 & 1.2 & \\
O & 8.7 & 1.8  && 8.7 & 1.5  && 7.5    & 0.5  && 8.9 & 1.5  && 8.6 & 2.2 && 8.9 & 2.2 & \\
Ne& ... & ...  && ... & ...  && 7.9    & 1.7  && ... & ...  && 9.6 & 4.0 && 7.9 & 1.9 & \\
Na& 5.9 & 1.6  && 5.8 & 1.2  && 5.9    & 1.5  && 6.4 & 1.6  && ... & ... && 5.5 & 1.3 & \\
Mg& 6.6 & 1.0  && ... & ...  && 6.1    & 0.4  && 6.2 & 0.1  && 7.3 & 2.2 && 6.0 & 0.6 & \\
Al& 5.3 & 0.8  && 5.4 & 0.6  && 5.6    & 1.0  && 5.7 & 0.7  && 5.9 & 1.9 && 4.7 & 0.2 & \\
Si& 7.6 & 2.0  && 7.3 & 1.4  && 7.5    & 1.8  && 7.0 & 0.9  && 8.1 & 3.1 && 6.3 & 0.9 & \\
P & ... & ...  && ... & ...  && 6.5    & 2.9  && ... & ...  && 5.8 & 2.8 && 4.2 & 0.8 & \\
S & 7.5 & 2.3  && 6.7 & 1.2  && 7.4    & 2.1  && 6.4 & 0.7  && 7.1 & 2.4 && 6.0 & 0.9 & \\
Ca& 5.1 & 0.7  && 5.0 & 0.3  && 5.3    & 0.8  && 5.1 & 0.2  && ... & ... && 4.2 &$-$0.1& \\
Sc& 2.8 & 1.6  && ... & ...  && 2.8    & 1.5  && 2.9 & 1.2  && ... & ... && 2.1 & 1.1 & \\
Ti& 3.3 & 0.3  && ... & ...  && 3.5    & 0.4  && 4.1 & 0.6  && ... & ... && 3.2 & 0.3 & \\
Cr& ... & ...  && ... & ...  && ...    & ...  && 4.2 & ...  && ... & ... && 3.7 & 0.1 & \\
Mn& ... & ...  && ... & ...  && ...    & ...  && ... & ...  && ... & ... && 4.3 & 0.9 & \\
Fe& 5.5 & 0.0  && 5.8 & 0.0  && 5.6    & 0.0  && 6.0 & 0.0  && 5.0 & 0.0 && 5.4 & 0.0 & \\
Ni& 5.6 & 1.4  && 5.2 & 0.6  && 5.8    & 1.5  && 5.9 & 1.2  && ... & ... && ... & ... & \\
Zn& 2.9 & 0.3  && 3.9 & 1.0  && 4.1    & 1.4  && 4.4 & 1.3  && ... & ... && ... & ... & \\
Sr& ... & ...  && ... & ...  && ...    & ...  && 2.2 & 0.7  && ... & ... && 0.5 &$-$0.3& \\
Y & 0.6 & 0.4  && 2.8 & 2.3  && 1.3    & 1.0  && 2.2 & 1.5  && ... & ... && ... & ... & \\
Zr& ... & ...  && 2.6 & 1.7  && ...    & ...  && 2.1 & 1.0  && ... & ... && 0.8 & 0.3 & \\
Ba& 0.7 & 0.6  && 1.4 & 1.0  && 0.9    & 0.7  && 1.3 & 0.7  && ... & ... && 0.5 & 0.4 & \\
\hline
\end{tabular}
$^a$abun = log $\epsilon$($el$)
\end{minipage}
\end{table*}

\begin{table*}
\begin{center}
\caption{The mean abundance ratios
$el$/S for cool EHe (including FQ Aqr), hot EHe (excluding V652 Her and
HD 144941) and majority class \rcb stars. The dispersion $\sigma$
is given in parentheses.}
\begin{tabular}{rrrrrc} \hline
\multicolumn{1}{c}{Ratio} & \multicolumn{1}{c}{hot EHe}
& \multicolumn{1}{c}{cool EHe} & \multicolumn{1}{c}{majority \rcb} & \multicolumn{2}{c}{Normal Star}\\
\cline{5-6}
& & & & \multicolumn{1}{c}{Sun}
& \multicolumn{1}{c}{[Fe/H] = $-$1} \\ \hline
Na/S& ... & $-$0.6 (0.2) & $-$0.8 & $-$0.9 & $-$0.6\\
Mg/S& 0.5 (0.2) & $-$0.1 (0.3) & ... & 0.4 & 0.4 \\
Al/S& $-$1.1 (0.3)  & $-$0.9 (0.3) & $-$1.0 (0.3) & $-$0.8 &$-$0.5\\
Si/S& 0.3 (0.3) & $-$0.2 (0.5) & 0.3 (0.2) & 0.3 & 0.3 \\
P/S & $-$1.4 (0.4) & $-$1.8 (0.3) & ... & $-$1.7 & $-$1.5:\\
Ar/S & $-$0.7 (0.1) & ... & ... & $-$0.7 & $-$0.7\\
Ca/S& $-$0.9 (0.1) & ... & $-$1.5 (0.3) & $-$0.9 & $-$0.9\\
Ti/S& $-$2.5 (0.5) & $-$2.6 (0.5) & $-$2.9 (0.2) & $-$2.2 & $-$2.2 \\
Fe/S& $-$0.1 (0.3) & $-$0.5 (0.4) & $-$0.4 (0.3) & 0.3 & $-$0.1\\
Ni/S&  ...     & ...  & $-$1.0 (0.3) & $-$1.0 & $-$1.4 \\
Zn/S&  ...     & ...  & $-$2.6 (0.3) & $-$2.6 & $-$3.0 \\
\hline
\end{tabular}
\end{center}
\end{table*}

\subsection{The C/He Ratios}

With two exceptions, the C/He ratios of the hot EHes are in the range
0.3\% to 1.0 \% for a mean of 0.7\%. Our cool EHes including FQ Aqr
 have C/He ratios
in the same range, as does the hot R CrB DY\,Cen.
The hot EHes with a  C/He ratio quite different from the mean are
V652\,Her and HD\,144941 with C/He $\simeq 0.003$\%. This pair also
show other differences with respect to the hot EHes.
The hot R CrB  MV Sgr has a  low ratio of C/He $\simeq$ 0.02\%.  
 The C/He ratio is not directly determinable for
the cool R CrB stars.

It is a remarkable result that, except for three stars with a very low C/He
ratio, the C/He ratio is uniform to within a factor of 3 despite large
variations in other elemental abundances affected by nuclear burning of H and He.

\subsection{Metallicity}

Products of hydrogen and helium burning are clearly present in the atmospheres
of the H-poor stars. To assess the initial metallicity of the stars, it is
necessary to consider elements unaffected by these
burning processes. Synthesis of elements from silicon through to the
iron group occurs in  advanced burning stages  and is followed by an
explosion.
Hence, we assume that their abundance
expressed as a mass fraction is preserved
and so indicates
a star's initial metallicity. There are three caveats:
(i)   synthesis of  these elements may occur in a companion star that
explodes and dumps debris on to the star that is or becomes the H-poor
star, (ii) severe exposure to neutrons will convert iron-peak and lighter
nuclei to heavier elements such as Y and Ba,
and (iii) the abundances may be distorted by non-nuclear processes.
Israelian et al. (1999) propose
the first of these scenarios to account for the composition of the
main sequence companion 
 to a low-mass X-ray binary, but in this case the
affected star is not demonstrably H-poor. Conversion of Fe to heavy elements
would create enormous overabundances of the heavy elements. Giridhar, Lambert, \& Gonzalez (2000)
discuss how dust-gas separation has affected the atmospheric composition of 
certain RV\,Tauri variables. Gravitational settling and radiative
diffusion are other processes that can distort chemical compositions.

Of the potential indicators of a star's initial metallicity, Si, Ca, Ti, and
Fe have been measured across the EHes and R CrBs (see Table 3). Table 5
shows that abundance ratios with respect to S are normal  across
the groups with the following exceptions: Ca/S, Ti/S, and Fe/S are
0.3 to 0.7 dex lower for the majority R CrBs than expected for normal stars
of any metallicity. To within the dispersions, the same ratios for the
EHes are at their expected values. The R CrBs also show a higher Ni/Fe
ratio than expected. In light of the Ca, Ti, Fe, and Ni abundances for the
R CrBs, we elect to identify Si and S as providing the initial
metallicity ([M/H] where M$\equiv$ Fe)
 which we find from relations between [Si/Fe] or [S/Fe] 
and [Fe/H] for normal stars. The mean [Fe/H] obtained from these
two relations for a given Si and S abundance is represented as [M/H].
Note that the Si/S ratio is normal across the entire sample; the apparently
low Si/S ratio of cool EHes has a large dispersion.
 Relative to M, the majority R CrBs are 
Ca, Ti, and Fe deficient but Ni rich. These `anomalies' may reflect either
systematic errors in the abundance analyses of (presumably) the R CrBs, 
and/or real differences between the EHes and the R CrBs.
The fact that the
 Si/Fe and S/Fe ratios are increased greatly for the minority R CrBs
and FQ Aqr (Table 4) suggests that systematic errors are not the
sole explanation for the higher ratios of the majority R CrBs; 
physical parameters of Table 4's denizens  overlap those of the
majority R CrBs and the EHes. Figure 12 shows the [Si/Fe] versus [S/Fe].
It is difficult to suppress the hunch that the Fe abundance has been
altered for stars in Table 4, and possibly for some stars contributing
to Table 3.
Other elements including Si and S may have
been affected to a lesser degree.
Adopting M as the initial metallicity, we show in Figure 13 histograms
for EHes and R CrBs; the 
mean values of M are slightly sub-solar.

\begin{figure}
\epsfxsize=8truecm
\epsffile{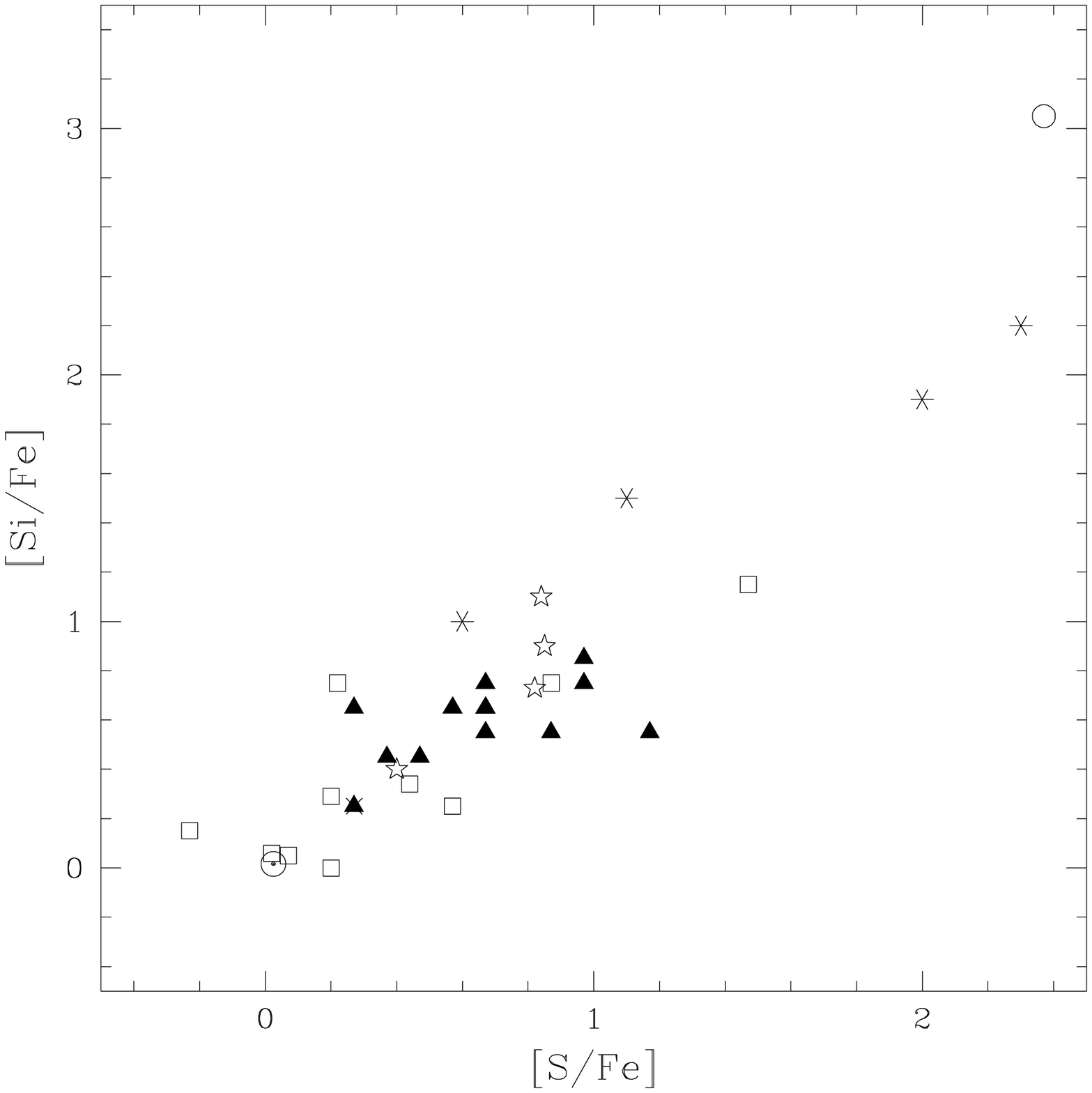}
\caption{[Si/Fe] versus [S/Fe] for cool EHe, majority and minority class \rcb and hot EHe stars: 
Symbols open $\star$ represent cool EHe, $\sq$ hot EHe stars, solid $\triangle$ majority class R~CrB, 
$\ast$ minority class R~CrB, $\bigcirc$ DY Cen, and
$\times$ V652 Her with low C/He ($\sim$ 0.003\%). The Sun is denoted by $\odot$.}
\end{figure}

\begin{figure}
\epsfxsize=8truecm
\epsffile{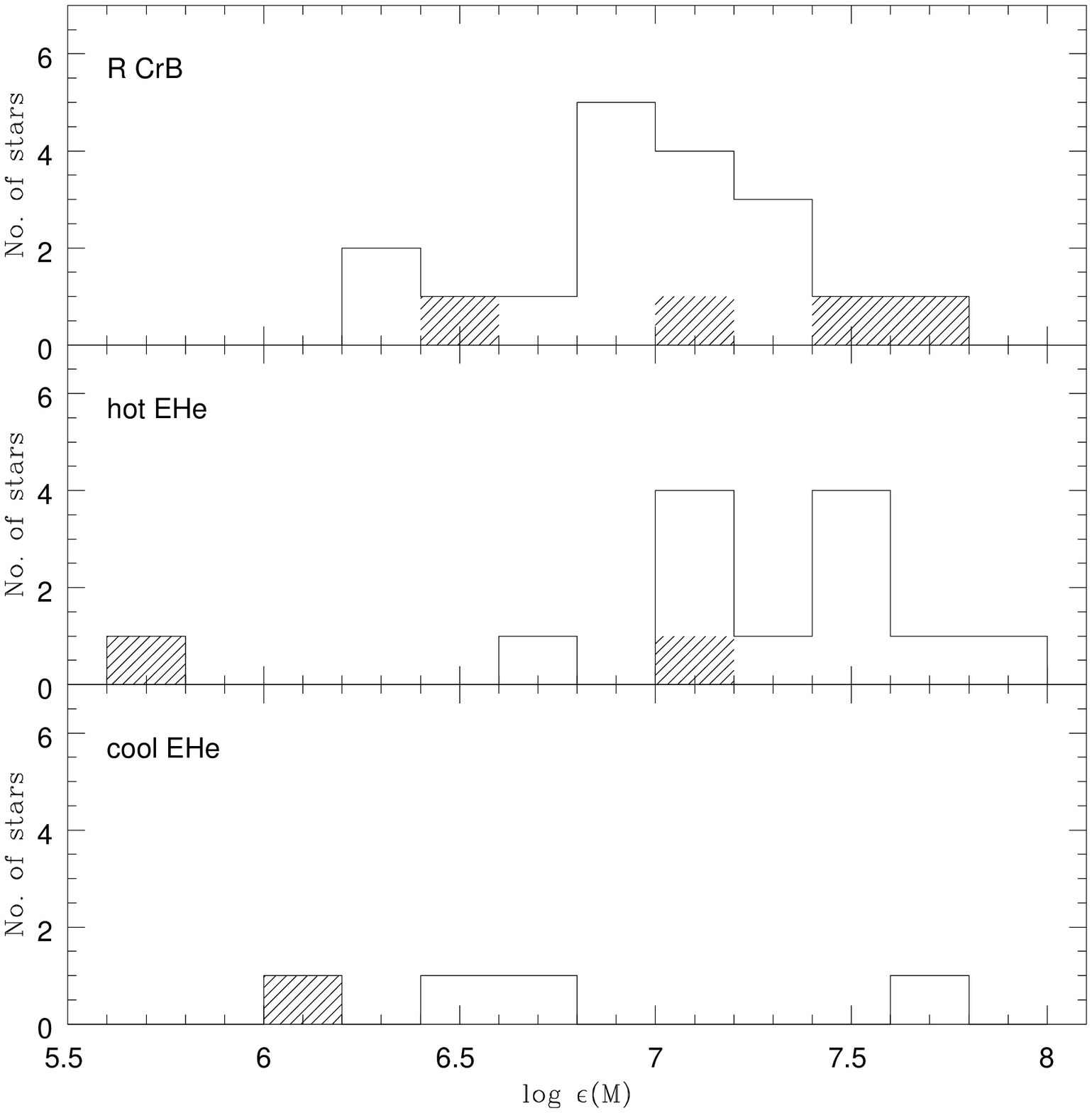}
\caption{A frequency histogram of the initial metallicity M for R~CrB, hot EHe and cool EHe stars.
The hatched stars represent the minority class R~CrBs, the cool EHe FQ~Aqr, the hot R~CrB DY~Cen, and the hot EHe HD\,144941 with 
low C/He content ($\sim$ 0.003\%).}
\end{figure}

To confirm the slight metal deficiency, we consider the star's Galactic distribution.
We have estimated
the mean vertical height $\langle|z|\rangle$ and space motions $(U,V,W)$
for 15 R CrBs. 
Distances were estimated assuming $M_V = -4.5 \pm 0.5$
and E(B-V) values were taken from Asplund et al. (1997a).
Proper motions and radial velocities were obtained from SIMBAD, and
Asplund et al. (2000), respectively. For the R CrBs, we find
$\langle|z|\rangle$ $\simeq 550$ pc with the high velocity and high $z$
star UX Ant excluded. The mean Galactic rotation velocity is $\langle V_{rot} \rangle$
is about 180$^{+20}_{-60}$ km s$^{-1}$ which according to Chiba \& Yoshii (1998) implies
[Fe/H] $\simeq -0.6^{+0.3}_{-0.2}$ or log $\epsilon (\rm M)$ $\simeq$ 6.9$^{+0.3}_{-0.2}$, a value roughly 
consistent with M from Figure 13 within errors. Two
exceptions, UX Ant and VZ Sgr, are high-velocity stars and not
surprisingly are quite metal-poor. 

Distances and the $\langle|z|\rangle$ for most EHes were taken from
Heber \& Sch\"{o}nberner (1981). For stars not in their list, we
assumed $\log L/L_\odot = 4.1$ to calculate distances. Radial velocities
were taken from Jeffrey, Drilling \& Heber (1987). We did not consider LSS 3184, 
LS IV +6$^\circ$002, HD 144941 and V652 Her because they have
lower $L/M$ values. The Galactic rotation
velocity of 160$^{+40}_{-100}$ km s$^{-1}$ implies,
log $\epsilon (\rm M)$ $\simeq$ 6.7$^{+0.5}_{-0.3}$, a value roughly consistent with M from Figure 13.
  
In the discussion of the abundances, we adopt M ($\equiv$ Fe) derived from the Si and S
abundances, using [Si/Fe] and [S/Fe] versus [Fe/H] relations for normal stars, 
as the primary measure of a star's initial metallicity, but note in
several places, how adoption of the Fe abundances affects our conclusions.

\subsection{Hydrogen}

Of the chemical elements, hydrogen shows the greatest abundance
variation across the EHe - R CrB sample: the most H-rich star is the hot R CrB
DY~Cen with log $\epsilon (\rm H)$ = 10.7 and the least H-rich stars are the R CrB stars
XX Cam and the minority member V3795~Sgr with log $\epsilon (\rm H) \leq 4$.
 For the majority R CrB stars,
the spread is from 7.4 for SU Tau to the upper limit of 4.1 for XX~Cam.
The hot EHes have a higher maximum H abundance: five of the six stars
with detectable Balmer lines have an abundance higher than that of the majority
R CrB stars. The lower bounds on the H abundance may not differ:
the upper limit to the H abundance
is comparable to the abundance in SU~Tau for four hot EHes.
The two hot EHes (V652\,Her and HD\,144941) with a very low C/He ratio are exceptionally H-rich.
MV~Sgr, the hot R CrB with a low C/He ratio, shows Balmer lines but its H
abundance has not been determined.

For HD\,144941, the EHe with the low
C/He ratio, and a hydrogen abundance log $\epsilon (\rm H) = 10.3$ (Harrison \&
Jeffery 1997), the measured H abundance offers a direct clue to the
evolution. If the hydrogen represents a residue of original material
that has not been exposed to H-burning, the {\it minimum} abundance
for heavier elements is readily predicted from the solar abundances
scaled to the H abundance. This assumes, of course, that the original
material was approximately a solar mix of elements. The reported abundances
for C to Fe are  within 0.3 dex (i.e., the errors of measurement)
 equal to the
adjusted solar values. There appear to be two 
extreme interpretations of this result. The low metallicity of HD\,144941
may be the initial metallicity but this is weakly contradicted by the
relative abundances that do not show the enhanced Mg/Fe and Si/Fe
ratios expected of such a metal-poor star. Or if the hydrogen was accompanied by
a solar mix of elements, the He-rich material with which it mixed was
very metal-poor. It is also the case, as noted by Harrison \& Jeffery (1997),
that not only is the C abundance low but there is no indication of
the N enrichment expected for He-rich material. We suggest that
H-rich material of solar metallicity was mixed with He-rich material
that had experienced gravitational settling of heavier elements.

For all but one of the  other stars (including V652\,Her),
the measured Fe/H ratio 
 is 2 or more orders of magnitudes greater than the solar
ratio; no useful information is thus provided by the H abundance about the
metallicity. 
 DY Cen is a curious exception. 
It is relatively
H-rich (Table 4). If this hydrogen is unburnt material of solar composition,
the Fe abundance expected is log $\epsilon (\rm Fe)$ = 6.3. If the hydrogen in this
material has been partially consumed, this is a lower bound to the expected
Fe abundance. The measured Fe abundance is 1.3 dex below this limit,
indicating that either the star was initially metal-poor by at least 1.3
dex or the H-rich material has a peculiar composition. Jeffery \& Heber
(1993) did comment on dust-gas separation. 
Unlike the case of HD\,144941, other elements do not mirror Fe. In fact, the
measured abundances for elements other than Fe
 exceed the H-adjusted abundances by 0.6 dex or more,
or, as shown in Figure 12, the S/Fe and Si/Fe ratios are extraordinarily
high. V854\,Cen (Table 4) may present a milder form of the puzzle offered
by DY\,Cen.

\subsection{Nitrogen, and Oxygen}

These elements with carbon
provide the principal record of the nucleosynthesis and
evolution that produced these H-poor stars. Nitrogen enrichment is the
signature of H-burning CNO-cycled products. Carbon is most
probably the leading product of He-burning. Oxygen and neon may accompany the
carbon. Although one might suppose
H-burning products to enter the atmosphere before the He-burning products,
one should consider the possibility that nuclear processing may continue after
the two products have mixed; possibly, H-burning
continues or resumes after He-burning products enter the envelope. One
hopes that the C, N, and O abundances will reveal what happened.
Normally, one would consider ratios N/Fe and O/Fe to be more accurate
reflections of the N and O changes because not only  are 
certain errors minimized when using an abundance ratio but also
the ratios allow for the changes of initial N and O
abundance with initial Fe abundance. But in Sec. 6.2 we suggested
that the high Si/Fe and S/Fe ratios of key stars implied a depletion
of Fe, and suggested using the mean of Si and S abundances.

\begin{figure}
\epsfxsize=8truecm
\epsffile{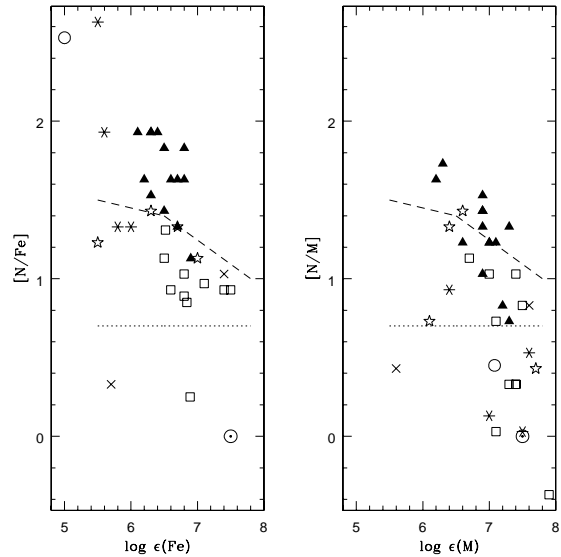}
\caption{[N/Fe] versus log $\epsilon (\rm Fe)$ and [N/M] versus log $\epsilon (\rm M)$
for cool EHe, majority and minority class \rcb and hot EHe stars:
Symbols open $\star$ represent cool EHe, $\sq$ hot EHe stars, solid $\triangle$ majority class R~CrB, $\ast$ minority class R~CrB, 
$\bigcirc$ DY Cen, and $\times$ hot EHes with low C/He ($\sim$ 0.003\%). The Sun is denoted by $\odot$.
The dotted line represents conversion of the initial sum of C and N to N. The dashed line represents
the locus of the sum of initial C, N, O converted to N.}
\end{figure}

If N-enrichment is due solely to the CN-cycle, the resultant N
abundance is effectively the sum of the initial C and N abundances; the
C/N ratio at equilibrium for the CN-cycle is small. In the event that
the ON-cycle has acted, the N abundance is effectively the sum of the
initial C,N, and O abundances; the O/N ratio at equilibrium 
is small. Then, to interpret the N abundances, we need the
initial C, N, and O abundances as a function of metallicity.

The assumed initial C, and O abundances
are taken from Carretta, Gratton, \& Sneden (2000), that is
[C/Fe] = 0, and [O/Fe] = +0.5 for [Fe/H] $<$ --1 with a linear transition
to this value from [O/Fe] = 0 at [Fe/H] = 0. Additionally, we
take [N/Fe] = 0  but this is not a critical choice in view of the
larger abundances of C and especially O.
With these initial abundances, conversion of C to N, and C and O to N
leads to the predicted trends shown in both
panels of Figure 14. Presently, there is
a debate over the true initial O abundances. Analyses of ultraviolet
OH lines (Israelian, Garc\'{i}a L\'{o}pez, \& Rebolo 1998;
Boesgaard et al. 1999)
 show [O/Fe] to increase linearly with decreasing [Fe/H] attaining
[O/Fe] = +0.8 at [Fe/H] = --2.0 or $\log \epsilon$(Fe) = 5.5, the
limiting inferred initial metallicity of our sample. If the OH-based trend
were correct, conversion of C and O to N at low metallicity continues
the upward trend shown in Figure 14 for metallicities greater than
$\log \epsilon$(Fe or M) = 5.5. Evidence is accummulating, however,
that the high [O/Fe]
values are likely overestimated (Nissen, Primas, \& Asplund 2000;
King 2000; Lambert 2000; Asplund 2000; Asplund \& Garc\'{\i}a P{\'e}rez 2001).

Observed N and Fe or M (from Si and S)
 abundances are compared in Figure 14 with the
predictions for CN- and ON-cycling based on the initial C, N, and O
abundances. The key points that may be made include
the following where unless
indicated otherwise M not Fe is assumed to provide the initial C,N, and O
mix:

First, the majority
R CrBs have  a maximum  N abundance equal to that  predicted
from conversion of initial C and O to N. The few exceptions have
a N abundance between that predicted from conversion of
C to N and C and O to N. The Fe abundance implies lower initial
C,N, and O such that the observed N abundances of most R CrBs
exceed the prediction for conversion of initial C and O to N.

Second, the N abundances
of the EHes are generally
 clustered between the predictions for N from CN- and
ON-cycling and appear to form an extension of the trend presented
by the majority R CrBs. Adoption of Fe
 as the metallicity indicator compresses the distribution of R CrBs and EHes.
Consistent with the idea that the N is a product of H-burning is the
fact (Sec. 6.3) that the H abundance of the R CrBs is on average
less than that of the EHes. 

Third, just three stars fall significantly off the
N/M vs M trend with a lower M than suggested by their N/M ratio:  
the minority R CrB VZ Sgr,
the cool EHe (FQ Aqr), and the
low C/He EHe HD\,144941. These stars provide the impression that there may
be a distinct subclass of He-rich stars.

Fourth, although the N abundances imply wholesale conversion of O to N
via the ON-cycle, many stars are not O-deficient (Figure 15).
Operation of the CNO-cycles reduces the C and O abundances. Obviously, the
high C abundance of the R CrBs and EHes implies C production from
He-burning. At ON-cycle equilibrium, the O abundance is about 1 dex
at 20 million K to 2 dex at 50 million K below its initial value.
In a few R CrBs and EHes, the observed O abundance is 1 dex below the inferred
initial abundance. In other stars, the O abundance may be as much 1 dex 
overabundant relative to the initial abundance. This implies (see below)
that O was synthesized along with the C, i.e., the 3$\alpha$-process
was followed by $^{12}$C$(\alpha,\gamma)^{16}$O.
The most O-rich stars have an observed  O/C ratio of
near unity implying roughly equal production of the two elements from 
He-burning.


\begin{figure}
\epsfxsize=8truecm
\epsffile{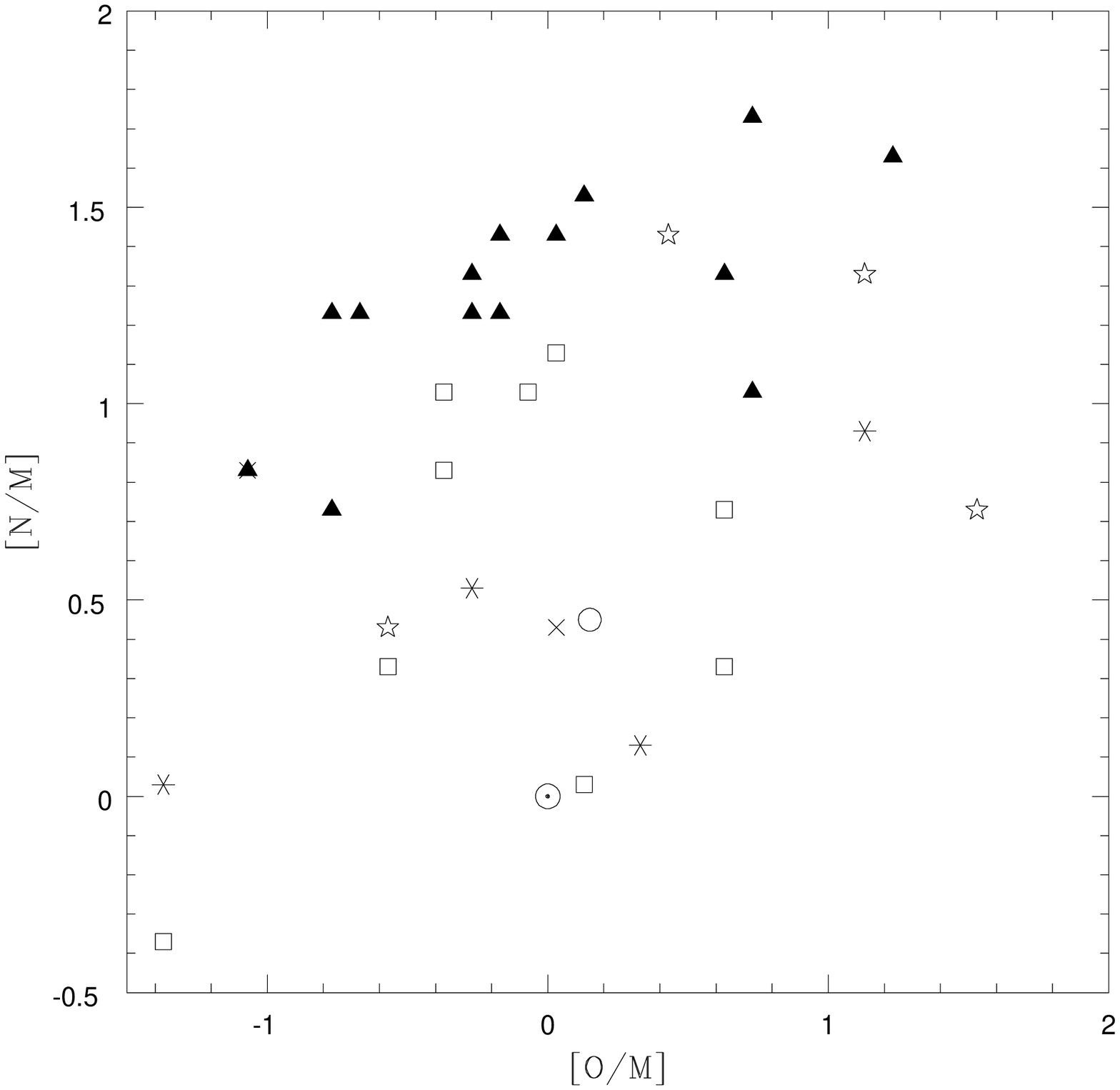}
\caption{[N/M] versus [O/M]
for cool EHe, majority and minority class \rcb and hot EHe stars:
Symbols open $\star$ represent cool EHe, $\sq$ hot EHe stars, solid $\triangle$ majority class R~CrB, $\ast$ minority class R~CrB,
$\bigcirc$ DY Cen, and $\times$ hot EHes with low C/He ($\sim$ 0.003\%). The Sun is denoted by $\odot$.}
\end{figure}


\subsection{Neon}

Neon abundances are known for four hot EHes including the low C/He star
V652\,Her, the four cool EHes, DY Cen, the minority R CrB V3795~Sgr, and the
majority R CrB Y Mus. In all but two stars neon is overabundant. ([Ne/M] $>$ 0).
The two stars of low Ne abundance
could be claimed to be unrepresentative of the family:
 V652~Her has a low C/He ratio for
a EHe star, and V3795~Sgr is a `peculiar' R CrB.
In the remaining stars, neon is grossly overabundant: [Ne/M] $\sim$
1 - 2.

Except for the hottest stars, the Ne abundances are based on Ne\,{\sc i} in
the red. Although the atmospheres are different, we recall that Auer
\& Mihalas (1973) in a classic paper on non-LTE effects found LTE Ne
abundances to be about 0.7 dex greater than the non-LTE abundances for
normal (i.e., H-rich) B stars with T$_{\rm eff}$ of 15000 to 22500 K
and $\log g$ = 3 to 4. If similar effects are present for this sample
of He-rich stars, our Ne abundances are overestimated. In the hottest EHes,
Ne\,{\sc ii} lines in the blue provide the Ne abundance (Jeffery 1996).

Here, we shall assume that Ne is substantially overabundant in at least
some of the EHes because it has been synthesized and added to the
atmosphere. Synthesis could have occurred from exposure of CNO-cycled material
to temperatures somewhat less than those required for He-burning.  In
such circumstances, $^{14}$N through two successive $\alpha$-captures
is converted to $^{22}$Ne. In He-burning at temperatures above about
250 $\times 10^6$ K, $^{22}$Ne$(\alpha,n)^{25}$Mg destroys the Ne
and serves as a neutron source.
An instructive way to
examine the data is to compare the sum of the N and Ne abundances as a function
of the initial metallicity, as in Figure 16. 
 
In this figure, the sum of the N and Ne abundances
 is shown relative to the predicted N
abundance resulting from thorough conversion of initial C and O to N.
The upper envelope of the points matches the predicted trend well.
This result suggests  an
explanation for the  generally
lower N abundances of the EHes relative to the R CrBs: the two groups 
experienced severe CNO-cycling resulting in conversion of 
initial C and O to N. But in the case of the EHes, substantial amounts
of the synthesized N was exposed to hot $\alpha$s and converted to Ne.
It will be necessary to extend the Ne measurements
to the R CrBs in order to show that Ne is not overabundant in them, a difficult
task given the lower temperatures of the stars.

In principle, Ne ($^{20}$Ne) may also be made by an extension of He-burning
by $\alpha$-capture on $^{16}$O at high temperatures.
Since the two modes of Ne synthesis produce
different isotopes, we note the possibility of measuring the isotopic
ratio. Odintsov (1965) measured the $^{22}$Ne - $^{20}$Ne shifts for
a selection of red Ne\,{\sc i} lines finding values from 0.015 to
0.038 cm$^{-1}$. Since a  shift of 0.038 cm$^{-1}$ for
 a typical line is equivalent 
to a velocity shift of only 0.7 km s$^{-1}$, characterization of the
isotopic mix will be difficult even from high-resolution
high-S/N spectra of the sharpest-lined stars.



\begin{figure}
\epsfxsize=8truecm
\epsffile{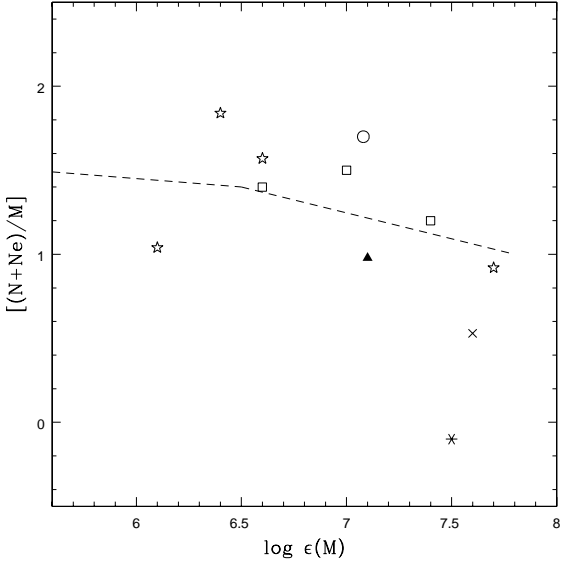}
\caption{[(N+Ne)/M] versus log $\epsilon (\rm M)$ for cool EHe, majority and minority class \rcb and hot EHe stars:
Symbols open $\star$ represent cool EHe, $\sq$ hot EHe stars, solid $\triangle$ majority class R~CrB,
$\ast$ minority class R~CrB, $\bigcirc$ DY Cen, and
$\times$ V652 Her with low C/He ($\sim$ 0.003\%). The dashed line represents the locus of the sum of initial C, N, O converted to N.}
\end{figure}


\subsection{Sodium to Calcium}

There is useful data on abundances in  both the EHe and
R CrB samples on Al, Si, and S. Data on P and Ar are  effectively
available only  for the EHes, on  Ca for a  few EHes but all R CrBs,
and Na only for the R CrBs.

The Al/S ratios
are close to the solar ratio and uniform across the cool and hot
EHe and majority
R CrBs. One EHe stands out with a low Al/S ratio: BD-9$^\circ$~4395,
the star with emission
lines: a sub-solar Al and supra-solar S abundance combine for a Al/S
ratio of --2.3 dex compared to the solar ratio of --0.8 dex. HD~168476 
appears to have a high Al/S ratio: Al/S = 0.2 dex, an
uncertain value from Walker \& Sch\"{o}nberner (1981) and much above the
value of --1.0 from Hill (1965). Neither discrepant value deserves great weight
at this time. Excluding these values, the mean abundance
ratio of Al/S is --1.1 $\pm$0.3 dex from 8 EHes which compares well with
the mean of --1.1 $\pm$0.2 dex from 12 majority R CrBs. 
The cool EHes appear to have a similar Al/S ratio.
The hint that the stellar ratios are less than the solar value of --0.8 dex
is compatible with a slightly sub-solar initial metallicity for the He-rich
 stars.
 There
is a hint too that the ratios may differ from the solar ratio of --0.8 dex in
the three minority R CrBs but not in V854~Cen,
Al is underabundant relative to S by about 1 dex. 

The  P/S ratio is solar within the
measurement errors for the EHes: the mean P/S ratio is --1.4 $\pm$0.4 and
--1.8 $\pm$0.3 dex for the hot and cool EHes, respectively.
Three hot EHes are suspected of a higher P abundance:
HD~168476, LSE~78, and LS~IV~+6$^\circ$~002 which are P-rich 
by 0.7 to 1.0 dex relative to the solar P/S ratio.
 For the cool EHe's BD~-1$^\circ$~3438,
and LS~IV~-1$^\circ$~002, the P/S ratio is normal.
 Among the R CrBs (Asplund et al.
2000), phosphorus was measured in  Y Mus (majority R CrB) and V3795 Sgr
(minority R CrB) and reported to be overabundant by about 1 dex. 
Additional spectroscopic scrutiny is needed before a conclusion is reached
concerning P in R CrB stars. 

Argon abundances for 7 hot EHes give a mean Ar/S ratio of
--0.7 $\pm$0.1 dex, i.e., the solar ratio. The only other measurement for our He-rich
stars is for the hot R CrB DY Cen which 
has Ar/S = --1.0 dex, i.e., solar within the measurement errors.

Calcium in four of the five hot EHes in which it has been measured
is of normal abundance: $\log[\epsilon$(Ca)/$\epsilon$(S)] $\simeq$ --0.9,
i.e., the ratio expected for solar and metal-poor stars. The exception is
HD~168476 for which Walker \& Sch\"{o}nberner (1981) give
Ca and S of equal abundance. Of the four cool EHes, two
have the  Ca/S (and/or  Ca/Si) ratio  close to normal ratio. The other
two, FQ~Aqr and LS~IV~-14$^\circ$~109, have Ca/S (and Ca/Si) about 1 dex
below normal. Among the majority R CrBs, the Ca/S  ratio is 0.6
below normal with very little star-to-star scatter. Among the minority R CrBs
the  Ca/S  ratio is lower by about an additional 0.6 dex and
similar to those of FQ~Aqr and LS~IV~-14$^\circ$~109. One interpretation of
these in Ca/S and Ca/Si ratios is that they reflect systematic errors that 
vary from hot to cool stars. Another speculation is that the low Ca
abundance is a signature of alteration of the compositions by a process
such as the winnowing of dust from gas.

Sodium abundances are unknown for the hot EHes. In the cool EHes, the
Na/S ratio is solar or slightly higher but the higher abundances are
in part or in whole based on a strong line in a region rich in telluric
H$_2$O lines. We give low weight to these results. In the majority R CrBs, the 
Na/S ratio is normal (i.e., solar) at --0.8 $\pm$0.3 dex.

\subsection{Heavy Elements - Y and Zr}

Our chief interest in the heavy elements is as tracers of
exposure to the $s$-process.
Unfortunately, heavy elements such as Y and Ba are difficult
to detect in the spectra of the EHes. Therefore, our data
are restricted to the R CrBs and a couple of the cool EHes. We emphasize the
data for Y and Ba, but Sr was measured for all cool EHes, and Zr and La
in some cool EHes, and  R~CrBs. To within the errors of measurement, the
Sr/Y, Zr/Y, and La/Ba
ratios are as expected for normal stars.

The abundance of Y and Ba expressed as [Y/M] and [Ba/M] vs [M] show
a scatter at a
given [M] (Figure 17 and 18).
 The mean [Y/M] is slightly positive but [Ba/M] scatters about zero.
The lighter $s$-process elements seem enhanced relative to the
heavy $s$-process elements.
This mild (Y) to
undetectable (Ba) $s$-process enrichment is at odds with an identification
of the stars as remnants of H-rich thermally pulsing AGB stars. For example,  
Ba stars typically show enhancements of 0.8 to 1.6 dex with a slight
increase in the Ba to Y ratio with decreasing metallicity. If He-rich
stars were previously on the AGB, they did not experience the third dredge-up.

These conclusions based primarily on $s$-process abundances measured for
the R CrBs are dependent on identifcation of M not Fe as the initial
metallicity. If Fe is chosen in preference to the Si-S abundances,
Y/Fe and Ba/Fe indicate a modest $s$-process enrichment in the R CrBs.

\begin{figure}
\epsfxsize=8truecm
\epsffile{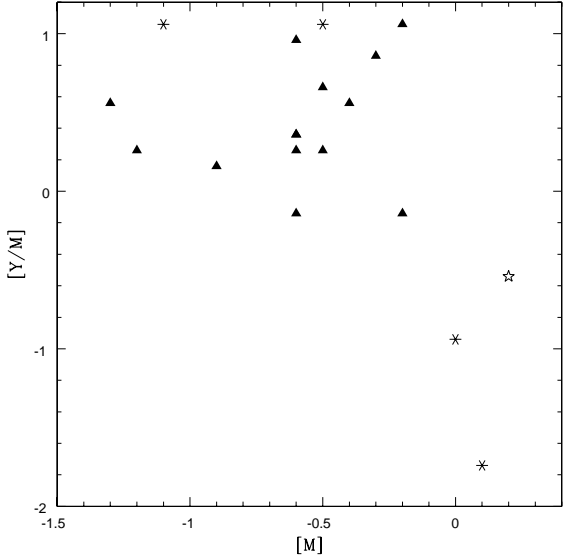}
\caption{[Y/M] versus [M]
for cool EHe, majority class \rcb and minority class \rcb stars:
Symbols open $\star$ represent cool EHe, solid $\triangle$ majority class R~CrB, and $\ast$ minority
class R~CrB.}
\end{figure}

\begin{figure}
\epsfxsize=8truecm
\epsffile{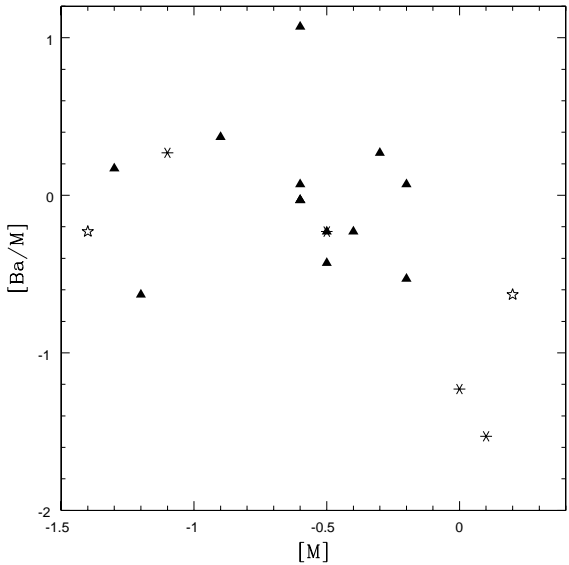}
\caption{[Ba/M] versus [M]
for cool EHe, majority class \rcb and minority class \rcb stars:
Symbols open $\star$ represent cool EHe, solid $\triangle$ majority class R~CrB, and $\ast$ minority
class R~CrB.}
\end{figure}


\section{A Simple Recipe and Application}

\subsection{Recipe}

The overall impression gained from the inspection of the spectra and
confirmed by the abundance analyses
is that the atmosphere of a H-deficient star most likely  consists of three
ingredients: a small fraction of H-rich material with original elemental
abundance ratios (except, perhaps, for some conversion of C to N), a large
fraction of  severely H-depleted He-rich material, and a fraction of
material from a He-burning layer. The three layers may not necessarily
have belonged to a single star. The principal tracers of the three layers
are H, N, and  C   with  O, respectively.  Our  simplest recipe  assumes that
there was no processing after mixing of ingredients. In particular, we assume
 that the C-rich material from the He-burning layer was not subsequently
exposed to H-burning CN-cycling. If we take the metallicity M derived from
the mean of Si and S, the maximum, observed N abundance may be accounted for
without a call for CN-cycling of the C-rich material from the He-burnt layer.

Perhaps, the single most striking aspect of the EHe's  is the near
 uniformity of their C/He ratios. For all, except HD\,144941 and V652\,Her,
 the ratio is in the narrow range
 0.3 -- 1.0 \%, a range not much greater than that expected
 from errors of measurement. 
 The C/He ratio of R CrB's is not directly
 measurable. The hot R CrB DY\,Cen has a ratio matching that of the
 majority EHe's but MV\,Sgr has a much lower
ratio. In its simplest form, our recipe provides 
 He from H-burnt material but C only from the He-burnt
material (in the exceptional case  of  HD\,144941, the surviving H
is accompanied by a significant amount of carbon). The observed C/He
 ratio, then, results from a mixing of a mass of H-burnt material $M_{CN}$
 with mass fractions $f_{CN}$(He)  $\simeq$ 1 and  $f_{CN}$(C) $\simeq $ 0, and a
 mass of He-burnt material $M_{3\alpha}$ with mass fractions
 $f_{3\alpha}$(He),  $f_{3\alpha}$(C), and $f_{3\alpha}$(O) for He,
 C, and O, respectively where $f_{3\alpha}$(He) + $f_{3\alpha}$(C) +
 $f_{3\alpha}$(O) = 1.
If  $m_{mix} = M_{3\alpha}/M_{CN}$,  the C to He ratio is given by

\begin{equation}
\frac{Z({\rm C})}{Z({\rm He})} = \frac{f_{3\alpha}({\rm C})m_{mix}}{1 + f_{3\alpha}({\rm He})m_{mix}}
\end{equation}

where $Z(X)$ is the mass fraction of element $X$ in the mixed envelope.

The near-uniformity of the C/He ratios allows for several
 possibilities.  Suppose He-burning ended far from
 completion: $f_{3\alpha}$(He) $\simeq$ 1 with
$f_{3\alpha}$(C) $\ll 1$. If, then, $m_{mix} \gg 1$,
 the observed C/He ratio is given by
$Z$(C)/$Z$(He) $\simeq f_{3\alpha}$(C).
This demands that all stars
 run He-burning to very similar and incomplete levels.
Alternatively,  He-burning ran to near-completion with
$f_{3\alpha}$(C) $\simeq 1$ (or, more properly,
  $f_{3\alpha}$(C)  + $f_{3\alpha}$(O)$
\simeq 1$ with  similar mixes of C and O, and even Ne in some cases).
 Now, $Z$(C)/$Z$(He)
 $\simeq f_{3\alpha}$(C)$m_{mix}$ with $m_{mix} \ll 1$,
 which seems a more palatable
 recipe in that moderate
variations of $f_{3\alpha}$(C) occasioned by temperatures and the timescale  of
He-burning are allowed.  It remains, however, unclear what constrains the
mixing fraction to small values.
 Oxygen production depends not only on the rate constant
 for $^{12}$C$(\alpha,\gamma)^{16}$O but also on the timescale
 (and temperature)  for  He-burning. If He is consumed quickly
 by the $3\alpha$-process, $^{12}$C is the principal product, but,
 if  He is consumed slowly, $^{12}$C may be converted to $^{16}$O.
  This means that the envelope's C/O ratio offers insight into
 how helium was burnt.

In our recipe,  the surface C abundance is not directly
dependent on the star's initial composition.
 This independence does not apply to N and O.
 Fortunately,  initial
C, N, and O abundances may be inferred (see above).
 Nitrogen  according to the recipe is
contributed  by the H-burnt material only.
  If H-burning is due to the CN- and ON-cycle,

\begin{equation}
Z({\rm N}) = \frac{Z({\rm C + N + O})_0}{1 + m_{mix}}
\end{equation}

which, if $m_{mix} \ll 1$, implies $Z$(N) $\simeq Z$(C + N + O)$_0$
 where the subscript $0$ denotes the star's initial composition.
If CN-cycling alone operates,
 $Z$(N) $\simeq Z$(C + N)$_0$.

According to our recipe, the O abundance in the limit that O is converted
to N by ON-cycling and $m_{mix} \ll 1$ is given by

\begin{equation}
Z({\rm O}) \simeq \frac{f_{3\alpha}({\rm O})m_{mix}}
    {1 + m_{mix}}
\end{equation}

ON-cycling is a slower process than CN-cycling. Calculations
(Arnould, Goriely, \& Jorissen 1999) show, however, that steady H-burning
results in a substantial reduction of
O   before H is completely converted to
He by the CN-cycle. This result applicable to a solar ratio
of C, N, and O to H will hold for reduced abundances of the
catalysts. Establishment
of the low O/N ($\sim$ 0.1) equilibrium ratio occurs for H deficiencies
of about  1 to 1.5 dex and greater for temperatures of 25 to 55
million K. At temperatures of 15 million K and less, consumption of O
is too slow and CN-cycling exhausts the H supply before O is depleted.
By contrast, CN-cycle participants are close to their
equilibrium abundances after consumption of just a few
protons per catalyst, that is a reduction of the H mass fraction by a mere
0.0004 from its initial value of about 0.75. These equilibrium abundances
persist as the ON-cycle participants attain equilibrium.
If CNO-cycling is the sole nuclear activity at temperatures of greater
than about 25 million K, one expects reduced C and O
and enhanced N abundances in material that is H-depleted by a factor
of about 1 dex or more, but a reduction restricted to the C abundance
alone in material that is H-depleted by a factor of {\it less} than
about 1 dex.  Given that the typical H-deficiency of the EHes and R CrBs
(Table 3) is 4 or more dex, O deficiencies (Figure 15) are
expected unless H-burning occurred at or below 15 million degrees K.
At ON-cycle equilibrium,
the O abundance is about 1 dex at 20 million K to 2 dex at 50 million K
below its value, assuming solar ratios, and  the N abundance is
greater than the surviving O abundance by corresponding amounts.

Inspection of Figure 15 shows that the
minimum O abundance is indeed about 1 dex below the initial value.
The N abundance for these stars is about 1.5 dex greater than
the O abundance, in fair agreement with CNO-cycle expectation.
Oxygen-deficient stars are, however, a minority among our stars.
EHes do not show the high N abundance expected
of O-deficient material.
Formally, our recipe allows for a reduced
N abundance (relative to that from CNO-cycling)
by increasing $m_{mix}$ in equation (4) but a large value
of $m_{mix}$ implies $Z({\rm O}) \simeq f_{3\alpha}({\rm O})$, and, in
all probability, an O abundance much greater than the observed range
[O/M] of -1 to +1.5. Thus, we do not favour this alternative.
A possible explanation is that the H-poor He-rich material
added to the envelopes resulted from H-burning at temperatures
of about 15 million K or less at which the ON-cycle is very slow.
This would preserve the O abundance but would not account for
the large spread in the O abundances (Figure 15), which would seem
to call  for addition of O from He-burning. The more likely explanation
suggested by the Ne abundances is that substantial amounts of N have been
converted to $^{22}$Ne.

Before proceeding further, we
recall the carbon problem presented by the R~CrBs.
Our conclusion about the greater degree of ON-cycled material in
R CrB stars is dependent on the
relative abundances of C, N, O, and the metallicity [M].
Recall that [M] generated from the mean of Si and S abundances
was chosen because Fe (also Ca and Ti) appears depleted, 
 especially for the R~CrBs and
quite apparently so for the minority R~CrBs. 
 What if [M] itself is not a fair
indicator of the initial metallicity of the R~CrBs?
For example, if [M] were increased by 0.5 dex on average, the
predicted N abundances following CN-cycling would be
similarly increased and would match the observed N
abundances.
Initial O abundances would be similarly increased by about 0.5 dex than the observed O
abundances.
Now,  an explanation not involving conversion of O to N
must be found for the O underabundances. What is needed, if
CN-cycling alone is to account for the atmospheric
compositions of R~CrBs, is a reduction of the N
abundances for almost all stars,
and an increase of the O abundances for at least the O-poor
stars. This is an unlikely combination
given the similarity of the dependence of the N\,{\sc i}
and O\,{\sc i} lines to the atmospheric parameters. Indeed,
most of the proposed solutions to the carbon problem
effectively preserved
elemental abundance ratios such as O/N (Asplund et al. 2000).
We suggest that it is likely that the \rcb atmospheres are
substantially contaminated with ON-cycled material.

\subsection{Application}


Dissection of the compositions of the majority \rcb stars is
summarized in Table 6 where stars are ordered by the difference
between the observed and predicted O abundance ($\Delta$ in Table 6). 
The observed N abundance is listed followed by the predicted abundance
assuming first that initial C is converted to N, and second that initial
O is also converted to N.
  Throughout the sample with the possible exception of RY~Sgr, 
the observed N abundances exceed the former prediction: the mean excess is 0.5$\pm$0.2 dex.
In contrast, the observed N abundances match the latter
 prediction: the mean difference
is 0.0$\pm$0.2 dex for all 14  stars.
 In our interpretation, material
present in these atmospheres is only heavily CN- and ON-cycled 
before He-burnt material was added without subsequent exposure to
hot protons.

 The O/C ratio of the He-burnt material
which is  estimated from the observed O and C abundances might be regarded
as an upper limit because we do not correct for surviving initial O and also the
C abundance used is the spectroscopic C abundance which is always lower than the
input C abundance for R CrBs.
The O/C ratio ranges from negligible to substantial, a range that may
be incompatible with assumptions behind our simple recipe.

\begin{table*}
\centering
\begin{minipage}{170mm}
\caption{Predicted and observed compositions of Majority R~CrBs where stars are ordered by the
difference between the observed and predicted oxygen abundance}
\begin{tabular}{lccrccccc} \hline
\multicolumn{1}{c}{}&\multicolumn{1}{c}{[M]} &\multicolumn{1}{l}{log $\epsilon (\rm O)_{obs}$}
&\multicolumn{1}{r}{${\Delta}^c$}
&\multicolumn{1}{c}{log $\epsilon (\rm N)_{obs}$}&\multicolumn{2}{c}{log $\epsilon (\rm N)_{predicted}$}
&\multicolumn{1}{c}{log $\epsilon (\rm C)_{obs}$}
&\multicolumn{1}{c}{O/C}\\
\multicolumn{1}{c}{}&\multicolumn{1}{c}{} &\multicolumn{1}{c}{} &\multicolumn{1}{r}{}
&\multicolumn{1}{c}{} &\multicolumn{1}{c}{log $\epsilon (\rm C+N)_{0}$} &\multicolumn{1}{c}{log $\epsilon (\rm C+N+O)_{0}$}
&\multicolumn{1}{c}{} &\multicolumn{1}{c}{(He-burnt)}\\ \hline
UV~Cas$^a$&--0.3 &7.5 &--1.1 &8.5 &8.4 &8.8 &9.2 &0.02\\
Y~Mus     &--0.4 &7.7 &--0.9 &8.8 &8.4 &8.8 &8.9 &0.06\\
FH~Sct    &--0.5 &7.7 &--0.8 &8.7 &8.3 &8.7 &8.8 &0.08\\
RY Sgr$^a$&--0.2 &7.9 &--0.8 &8.5 &8.6 &8.9 &8.9 &0.10\\
UW~Cen    &--0.9 &7.7 &--0.6 &8.3 &7.9 &8.5 &8.6 &0.13\\
V482~Cyg  &--0.6 &8.1 &--0.4 &8.8 &8.2 &8.7 &8.9 &0.15\\
GU~Sgr    &--0.5 &8.2 &--0.4 &8.7 &8.3 &8.8 &8.8 &0.25\\
RT~Nor    &--0.2 &8.4 &--0.3 &9.1 &8.6 &8.9 &8.9 &0.30\\
RS~Tel    &--0.6 &8.3 &--0.1 &8.8 &8.2 &8.6 &8.9 &0.30\\
XX~Cam$^b$&--0.6 &8.4 &0.0  &8.9 &8.2 &8.6 &9.0 &0.30\\
SU~Tau    &--1.2 &8.4 &0.3  &8.5 &7.7 &8.3 &8.8 &0.40\\
R~CrB$^b$ &--0.6 &9.0 &0.5  &8.4 &8.2 &8.7 &9.2 &0.60\\
RZ~Nor    &--0.6 &8.9 &0.5  &8.7 &8.1 &8.6 &8.9 &1.00\\
UX~Ant    &--1.3 &8.8 &0.8  &8.3 &7.6 &8.2 &8.9 &0.80\\
\hline
\end{tabular}
$^a$Incomplete CN-cycling?\\
$^b$CN-cycling of fresh C?\\
$^c$$\Delta$ = log $\frac{\epsilon (\rm O)_{obs}}{\epsilon (\rm O)_{0}}$
\end{minipage}
\end{table*}

Results for the hot and cool EHes (Table 7)
present a different picture.
For 5 stars, the observed N abundance is
appreciably (0.4 dex or more) less than predicted from
conversion of initial C to N. Except in two cases (BD -1$^\circ$3438
and LS IV -1$^\circ$002), the N abundance can be accounted for without
demanding ON-cycled material be now present in the atmospheres.
Recall that the available Ne abundances imply that the lower N
abundances of the EHes may be due to partial conversion of N
to $^{22}$Ne at temperatures just too cool for He-burning.
BD--9$^\circ$4395 whose spectrum shows emission lines is the only star in Table 7 with 
observed N less than the initial N; adoption of Fe as M however eliminates this problem.
The inferred O/C ratios for the He-burnt material span the
range inferred from the majority \rcb stars.
Note that V652\,Her and HD\,144941 with low C/He ($\sim$ 0.003\%) are not included in Table 7.

\begin{table*}
\centering
\begin{minipage}{170mm}
\caption{Predicted and observed compositions of EHes where stars are ordered by the
difference between the observed and predicted oxygen abundance}
\begin{tabular}{lrcrccccc} \hline
\multicolumn{1}{c}{}&\multicolumn{1}{c}{[M]} &\multicolumn{1}{l}{log $\epsilon (\rm O)_{obs}$}
&\multicolumn{1}{r}{${\Delta}^f$}
&\multicolumn{1}{c}{log $\epsilon (\rm N)_{obs}$}&\multicolumn{2}{c}{log $\epsilon (\rm N)_{predicted}$}
&\multicolumn{1}{c}{log $\epsilon (\rm C)_{obs}$}
&\multicolumn{1}{c}{O/C}\\
\multicolumn{1}{c}{}&\multicolumn{1}{c}{} &\multicolumn{1}{c}{} &\multicolumn{1}{r}{}
&\multicolumn{1}{c}{} &\multicolumn{1}{c}{log $\epsilon (\rm C+N)_{0}$} &\multicolumn{1}{c}{log $\epsilon (\rm C+N+O)_{0}$}
&\multicolumn{1}{c}{} &\multicolumn{1}{c}{(He-burnt)}\\ \hline
BD-9$^\circ$4395$^a$        &0.4 &7.9 &--1.4 &8.0 &9.1 &9.5 &9.10&0.05 \\
BD+10$^\circ$2179$^b$       &--0.2 &8.1 &--0.6 &8.1 &8.6 &8.9 &9.50&0.04\\
LS\,IV\,-14$^\circ$109$^e$  &0.2 &8.5 &--0.6 &8.6 &8.9 &9.3 &9.45 &0.10 \\
HD\,124448                  &0.0  &8.5 &--0.4 &8.8 &8.7 &9.1 &9.50&0.10\\
HD\,168476                  &--0.1 &8.4 &--0.4 &8.9 &8.7 &9.0 &9.50&0.10\\
LSS\,3184                   &--0.8 &8.1 &--0.3 &8.3 &8.0 &8.5 &9.00&0.10\\
LS\,IV\,+6$^\circ$002       &--0.5 &8.3 &--0.2 &8.5 &8.3 &8.7 &9.40&0.10\\
LSS\,99$^c$                 &--0.4 &8.6 &0.0 &7.6 &8.4 &8.8 &9.10 &0.30\\
BD-1$^\circ$3438            &--0.9 &8.4 &0.1 &8.5 &7.9 &8.5 &8.90&0.30\\
LSE\,78                     &--0.4 &9.1 &0.5 &8.3 &8.4 &8.8 &9.50&0.20\\
LS\,II\,+33$^\circ$005$^d$  &--0.1 &9.4 &0.6 &8.2 &8.6 &9.0 &9.40&0.70 \\
LSS\,4357$^d$               &--0.1 &9.4 &0.6 &8.2 &8.6 &9.0 &9.40&0.80 \\
LS\,IV\,-1$^\circ$002       &--1.1 &8.9 &0.8 &8.2 &7.8 &8.3 &9.30&0.30\\
FQ~Aqr                      &--1.4 &9.0 &1.2 &7.3 &7.5 &8.0 &9.20&0.60\\
\hline
\end{tabular}
$^a$Star has non-solar mix of metals. At [M] = 0.4, which seems extraordinarily high, observed N
is less than initial N. If Fe taken as initial metallicity ([M] = --0.9), N is consistent with CN-cycling prediction.\\
$^b$N is less than predicted by CN-cycling of initial C. Non-solar
  mix of metals. If Fe taken as initial metallicity ([M] = --1.0), N
  implies CN- and some ON-cycling.\\
$^c$No CN-cycling! Observed O very similar to inferred initial O and, therefore,
  O/C ratio of He-burnt material is very low.\\
$^d$Observed N less than expected from CN-cycling. [M] too high?\\
$^e$N abundance too low to account for the low O abundance.\\
$^f$$\Delta$ = log $\frac{\epsilon (\rm O)_{obs}}{\epsilon (\rm O)_{0}}$
\end{minipage}
\end{table*}

The  N abundances  of the
R~CrBs show that their atmospheres are
substantially contaminated by ON-cycled, as well as CN-cycled
material. The EHes have a lower N abundance that does not directly
imply substantial ON-cycling. When neon abundances are included in
the picture, it is seen that the sum of the N and Ne
abundances is equivalent to the N anticipated from the combination of CN-
and ON-cycling;  the N was apparently converted to $^{22}$Ne  by
 $\alpha$-captures prior to ignition of He.
This  difference between
EHes and R~CrBs has important
consequences for the understanding of the evolution of these
H-poor stars. Understanding can be deemed complete only
when the links, if any, between EHes and R~CrBs are identified,
and the progenitors and descendants of the stars are identified.
At present, the progenitors are unknown. Two scenarios are
discussed in Section 9.
Here, we
ask  - Do EHes evolve into R~CrBs? Or do R~CrBs evolve into EHes?
Or do the two classes have different origins? It is possible that
the answer is `Yes' to all three questions for subsets of the
stars.

Consider the case of evolution of EHes to R~CrBs. 
He-burning products are present in
the EHes such that C/He is uniform from one EHe to another with
a couple of exceptions. Given that C/N $\sim$ 10 for the EHes and a factor of
about 3 enrichment of N is needed to produce a R CrB, a relatively
mild processing of C to N is called for to turn a EHe into a R CrB.
This call for processing and mixing is reminiscent of the first dredge-up
in normal low mass red giants.
Noting that the O abundances of the two kinds of stars are similar,
CN-cycling must suffice for the production of additional N, a requirement
calling for low temperature H-burning. 
This picture must meet one additional constraint: the EHes must contain  
an adequate supply of protons to sustain conversion of C to N. Since the N abundance
must be increased from about $\log\epsilon$(N) $\sim$ 8 to 8.6 and each conversion
requires a minimum of 2 protons, the initial H abundance should exceed 8.8. This is close
to the maximum value seen in EHes. There is a rough anti-correlation
between high H abundance and low N abundance among EHes. 
Additionally, the R CrBs as a class have lower H abundances than the
EHes, and too little H to sustain significant additional conversion of
C and O to N.
The R~CrBs should  have a  lower C/He ratio than the
EHes. This prediction can not be excluded.
Evolution of EHes to R CrBs requires the R CrBs to have the high Ne abundances
seen in the EHes. The one measurement for a majority \rcb just meets this requirement.
Additional measurements of Ne in R CrBs are now needed. But a difficulty remains.
In this scheme, the N abundance of the R CrBs is the fruit of
 conversion of initial C and O to N supplemented in a major way
by conversion of fresh C to N. Then, the fact that their N
abundances are close to the maximum expected from total conversion of
initial C and O to  N is an accident. This is, perhaps, the
leading objection to the proposal that EHes evolve to R CrBs with 
additional N production via conversion of fresh C to N. Adoption of Fe
rather than Si and S are the indicators of a R CrB's metallicity removes
this objection.

The uniform C/He ratio of the EHes, and the generally lower N of EHes  are
obstacles to accepting
that evolution proceeds from the R CrBs to the EHes. Although N can be reduced
and Ne enhanced
by deep mixing that adds material raised to or beyond the temperatures
of He-burning, it seems likely that C is added at this time and
a dispersion in C/He created. An astute reader will appreciate that
the uniform C/He ratio is unexplained by us and simply assumed to
result from formation of an EHe. Clearly, the uniformity unless it is
an aberration of small number statistics
implies a narrow range of progenitors and a single mode of
formation for EHes.

The third scenario considers the  EHes and R~CrBs to differ from
birth. Subsequent evolution may drive EHes to higher temperatures
and R CrBs to lower temperatures;
for example, R~CrBs are likely evolve to higher $T_{\rm eff}$ and so appear
among the EHes,
but if the evolution is rapid relative to the lifetime
of EHes formed directly, contamination of the EHe sample
with evolved R~CrBs could be small. Despite differences in composition,
there are considerable similarities between EHes and R~CrBs that
suggest that they may have a common parentage.



There remain the minority \rcb stars, the hot \rcb star DY Cen, and the
low C/He ratio hot EHe stars. A key problem with the minority \rcb stars
is identification of their initial composition; Si and S (Table 4) imply
modest deficiencies ([M] $\sim$ 0.1 to --0.4, except for V854 Cen) but 
iron (and other elements) indicate [Fe] $\sim$ --1.7 to --2.4. Given this
distortion of elemental ratios, one wonders if C, N, and O
were immune. It probably suffices to comment on one star and allow
the reader to check that the other stars will also not neatly fit
our recipe. Consider V3795 Sgr for which [M] $\sim$ 0 but [Fe] = --1.9.
The observed N abundance is solar (i.e., the presumed initial value) not the
0.7 dex greater value expected from CN-cycling. To compound the puzzle, O
is depleted by 1.4 dex. Except for V854\,Cen, the other 3 minority \rcb
stars exhibit a similar problem. Nitrogen in V854\,Cen implies substantial
CNO-cycling and a O/C ratio of the He-burnt material of about 0.2.
The peculiar abundance ratios such as Si/Fe and S/Fe are difficult
to understand in terms of nucleosynthesis, except perhaps as a $rp$-
(effectively $r\alpha$-) process enriching $\alpha$-nuclides up to Si and S
but not Ca. Given the presence of post-AGB stars and RV Tauri variables
with photospheres highly depleted in elements that condense easily into
dust grains, one is tempted to invoke this process for the minority
R~CrBs, and in a milder form for the majority R~CrBs, Asplund et al.
(2000) discussed this idea.
A difficulty is that
hardy grains in C-rich environments are those of SiC yet Si is not depleted.
Since the EHes as
a class do not show highly anomalous Si/Fe and S/Fe ratios, we presume that the process operated
in the R~CrBs themselves or in their progenitors. Certainly,
extended atmospheres with a propensity to form dust are
auspicious sites for winnowing of dust from gas. Binarity seems to
encourage the winnowing in post-AGB stars and possibly the RV Tauri
variables (Van Winckel, Waelkens, \& Waters 1995; Giridhar, Lambert, \& Gonzalez 2000). Rao et al. (1999) speculated that
R~CrB may be a binary.

Finally, there are the two EHe stars with a low C/He ratio. HD\,144941 was
discussed in Sec. 6.3. V652\,Her with [M] $\simeq 0.1$ seems to fit
the picture of an atmosphere with considerable amounts of CN- and ON-cycled
material.
The low C abundance implies a small fraction of He-burnt material of
indeterminate O/C ratio. The merger of two helium white dwarfs could
result in formation of EHe stars with a low C/He ratio and lower luminosity (Saio \& Jeffery 2000).

Our discussion assumes Ne to be a product of $\alpha$-captures on N.
Attribution of Ne overabundances to extended He-burning seems
questionable because the Ne overabundances
are found in stars with low O/C and high O/C
from the 3$\alpha$-processed material.
High C and high Ne, at least
in our simple view of the He-burning zone, are incompatible; the Ne-rich
material is necessarily C-poor.

\section{The Theoretical Hertzsprung-Russell Diagram}

For comparisons with theoretical evolutionary tracks, it is convenient
to consider observed and theoretical stars in the Kiel
diagram of $\log g$ versus log $T_{\rm eff}$ where stars are
represented by observationally determined quantities.
On the assumption that He-rich stars once created evolve continuously,
different classes of the stars will appear connected in the H-R
diagram. Particularly rapid evolutionary phases connecting slower
phases will likely appear as unpopulated gaps, especially given small
sample sizes.
In this diagram (Figure 19), the hot EHe stars, cool EHe stars and 
the  \rcb stars
 form a quasi-continuous sequence suggesting but not requiring an
 evolutionary link between them. The sequence is bounded by lines
of constant $L/M$ corresponding to $\log L/M \sim 3.75$ to 4.5; evolution
from the AGB to the top of the white dwarf cooling track occurs at  constant
$L$. Also shown is the horizontal branch (HB), and the Extreme horizontal branch (EHB) stars
populate the very blue edge of HB (Figure 19).

Before turning to theoretical ideas on how to populate the constant $L/M$ strip
with He-rich stars, we comment on putative
relatives of the EHes and R CrBs, that is stars at either
end of the strip having compositions like those of the EHes and
R CrBs.	

\begin{figure}
\epsfxsize=8truecm
\epsffile{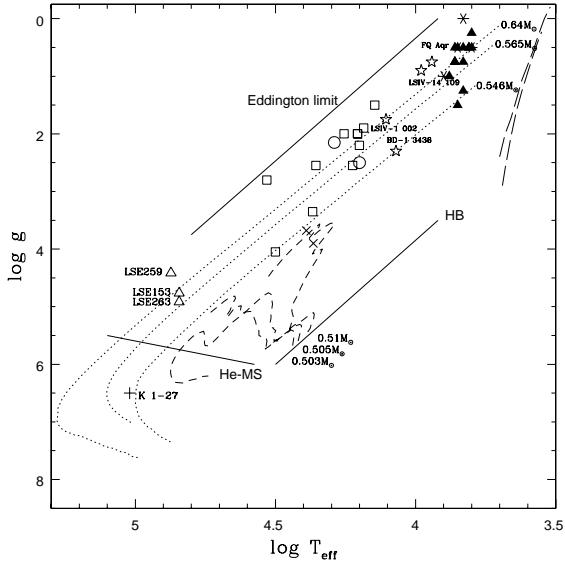}
\caption{$\log g$--log $T_{\rm eff}$ diagram for cool EHe, R~CrB, hot EHe, HesdO, and O(He) stars: 
Symbols open $\star$ represent cool EHe, solid $\triangle$ majority class R~CrB, $\ast$ minority class R~CrB, 
$\bigcirc$ hot R~CrB, $\sq$ hot EHe, $\times$ hot EHes with low C/He ($\sim$ 0.003\%),
$\triangle$ HesdO, and + O(He) stars.
Solid lines show the helium main-sequence, the horizontal branch and the Eddington limit for pure Thomson scattering
in a helium atmosphere. Evolutionary tracks of post-AGB (Sch\"{o}nberner 1983), post-EHB (Caloi 1989),
and RGB and AGB that begin from the HB (Giridhar, Lambert, \& Gonzalez 2000) are shown by dotted, short dashed, and long
dashed lines, respectively for different masses.}
\end{figure}

At the cool end are the cool hydrogen-deficient carbon stars (HdC) which
according to the only analysis available seem to be a close
match to the R CrBs (Warner 1967). Visible spectra of HdC are replete with
strong molecular lines that have proven a deterrent to quantitative
analysis.

Beyond the hot end of the EHe sequence are stars variously classified
as sdB, sdO including the central stars of planetary nebulae, and then
the white dwarfs. He-rich stars are found throughout this hot extension.
The sdO-sdB region is home to two kinds of stars. There are
stars of low $g$ that are called `luminous' sdB or sdO
stars, and at higher $g$ are `compact' sdB and sdO stars (Haas et al.
1996). The former are most probably post-AGB stars
and the latter post post-EHB stars; sample post-AGB and post-EHB evolutionary
tracks are shown in Figure 19.
Whether luminous or compact, the stars show
a wide range in H abundance and include He-rich objects.  In the
observed sample of luminous stars
 are some that qualify as putative relatives, presumably
descendants, of the EHes and R CrBs.

Husfeld et al. (1989) analysed 3 stars (LSE153, LSE259 and LSE263) with $T_{\rm eff} \sim 70000$K
and $\log g \simeq 4.7$ with no detectable H: the limit
H/He $\leq $ 5 to 10\% by number is not very restrictive because the
Balmer lines in these hot stars coincide with the Pickering lines of
He\,{\sc ii}. For 2 stars, C/He $\simeq$ 2\%, a value similar to that
of the EHes, but a third star is C-poor with C/He $\sim$ 0.01\%.
The N abundances are rather higher than found for EHes but at the 
upper end of the range for  R CrBs.
At higher effective
temperatures, He-rich stars exist among central stars of planetary
nebulae: e.g., the O(He) star in the planetary
 K 1-27 (Rauch, K\"{o}eppen, \& Werner 1994) with
$T_{\rm eff}$ = 10$^5$K and $\log g = 6.5$ has H/He $\leq 0.2$, a N
abundance slightly higher than that of a N-rich R CrB but no
detectable C or C/He $\leq 0.5$\%. This qualifies as a R CrB descendant. 
One supposes that the sequence of He-rich star ends in the DO white
dwarfs.
Although the paucity of data cannot be overlooked, it does seem
possible to identify relatives of the EHe and R CrBs 
between the hottest EHes and R CrBs and the start of the white dwarf
cooling track. Then, surface compositions may be altered by
gravitational settling such that they lose their memory of their
antecedents (Dreizler 1999).

There are other He-rich stars that judged by $T_{\rm eff}$ and $\log g$
could be relatives of the EHes and the R CrBs. These are the Wolf-Rayet
 [WC]-type central stars of planetary nebulae (CSPN) and the PG 1159 stars.
The He-rich CSPN have $T_{\rm eff}$ $\sim 75000$ to 180000K and
$\log g \simeq 5.5$ to 8.0. Their surface composition is typically
He:C:O $\sim$ 1:0.5:0.1 by number (Leuenhagen \& Hamann 1998), i.e., C and O are very
much more abundant than in the EHes and the R CrBs.  
PG 1159 stars have a similar composition. Dreizler \& Heber's (1998)
analyses of 9 stars give mean number ratios:
C/He = 0.3, O/He = 0.06 with very low N abundances (N/He $\leq 10^{-4}$ for
non-pulsators and 0.01 for the pulsators). Helium-burning products
are a more serious contaminant in these stars than in the EHes and R CrBs.
Evolution along a Post-AGB track will not
provide the necessary mixing. Then, it appears that these
C-rich stars and the EHe/R CrBs originate in different ways.

\section{Evolutionary Scenarios}

Recent discussions of EHes and R~CrBs have focussed on two
 theoretical proposals for their origins.
The Double-Degenerate (DD) scenario invokes the merger
of a He white dwarf with a C-O white dwarf. The final-flash (FF)
scenario supposes a late or final He-shell flash in a post-AGB
star that reexpands the star to giant dimensions.
Asplund et al. (2000) provided a discussion of the DD and FF scenarios and assessed their ability to
account for the compositions of R~CrBs. In a few cases, notably the low C/He EHes V652\,Her and HD\,144941,
a post-EHB origin may account for the stars.

Evolution off the AGB is very unlikely to produce
a He-rich star such as HdC or a R CrB directly by loss of the H-rich envelope.  
A H-rich post-AGB star may experience subsequently
a final He-shell flash, reconversion to an AGB-like star, destruction of 
hydrogen in its envelope, and a decline along a post-AGB track back to the
white dwarf cooling track.
Early discussions of the FF scenario are by Fujimoto (1977) and
Sch\"{o}nberner (1979), and Iben \& Renzini (1983). 
More detailed discussions by Herwig, Bl\"{o}cker, \& Driebe (1999)
and Herwig (1999) have shown how diversified the
FF scenario may be.

There is a fundamental problem in identifying EHes and R~CrBs
as post-AGB stars from a FF. The He-shell of an AGB star is rich in C and O
with mass fractions (He $\sim$ C $\sim$ O) if convective overshoot
is considered (Herwig 2000). Without overshoot the ratios are 
He:C:O $\approx$ 0.7:0.3:0.01 by mass fractions. The final He-shell
flash cannot greatly modify this mixture because the mass of the
H-rich envelope is necessarily very small. The observed ratios
are very different, for example, He accounts for more than 95\% 
of the He mass of an EHe. This sharp difference between the predicted
and observed compositions suggests that the EHe-R~CrB sample cannot
be comprised of post-AGB stars. These post-AGB stars should also
be rich in $^{13}$C but the $^{13}$C abundance is low according to
upper limits for \rcb stars. Absence of observed
$s$-process enrichments also rules out an association with thermally-pulsing
AGB stars.

Herwig (1999)  points out that the fate of a  star evolving directly
 off the AGB 
depends in large part on how recently it experienced a He-shell
flash or thermal pulse (TP). He identifies four principal
possibilities:
\begin{itemize}
\item
{\bf No additional TP.} If the star departs soon after a TP, it is
most likely to settle on the white dwarf track and not experience
another TP. Such stars are not progenitors of EHe and
R CrB stars.  
\item
{\bf A very late TP.} A more delayed departure from the AGB results in
a return from the white dwarf cooling track to the domain of the
EHe and \rcb stars. The H-rich envelope is engulfed by the He-burning
layers of the TP. Although H is thoroughly depleted by the models, the
problem is that
He, C, and O are produced with roughly equal mass fractions, and N is
not produced in significant amounts.
In addition, the depletion of hydrogen takes place by the reaction $^{12}$C(p,$\gamma$)$^{13}$C($\alpha$,n)$^{16}$O.
The produced neutrons are captured by
Fe seed nuclei resulting in enrichment of $s$-process elements but
the $s$-process elements in R~CrBs and cool EHes are solar like except for mild enhancement of
light $s$-process elements for majority class R~CrBs.

 Lithium is brought to the surface by dredge-up in a very late TP. The reaction
$^3$He($\alpha$,$\gamma$)$^7$Be(e$^+$$\nu$)$^7$Li, produces $^7$Li from $^3$He. $^3$He
enters into He-flash convective zone during a very late TP together with hydrogen from the
envelope. Lithium is present only in four R~CrBs out of the eighteen analyzed \rcb stars (Asplund et al. 2000).
The hot \rcb MV~Sgr shows Li in emission (Pandey et al. 1996).

\item
{\bf A late TP.} An even more delayed departure from the AGB leads to
a TP occurring before the H-burning shell has turned off. Subsequent
evolution follows that of the VLTP except the envelope is not mixed into
the He-shell, but dredge-up in the freshly renovated AGB star does
subsequently reduce the surface H abundance. In contrast to the VLTP,
the surface H abundance is reduced by a factor of only about 50.
The final He, C, and O (and N) surface mass fractions are
also unlike the observed abundances
of the EHe and the \rcb stars. 
\item
{\bf A very early TP.} Departure from the AGB immediately following
a TP occurs when the envelope mass is very small, and a dredge-up
commenced at this time reduces the H mass fraction by perhaps 20\%
at most.   
\end{itemize}

As argued by Sch\"onberner (1996), the FF scenario cannot account for
the EHes and R CrBs with their low C/He ratio. There are stars
that do match the predicted FF abundances quite well. In particular,
a rare few have evidently evolved from a WD dwarf-like state
to an AGB in a very short time, as the model predicts.
Sakurai's object (V4334\,Sgr) is most likely experiencing
a very late TP. Its surface composition evolved considerably
over a few months before the photosphere was obscured by
a thick cloud of dust (Asplund et al. 1997b, 1998, 1999). Herwig
(1999) showed that rapid changes of surface
composition are a feature of such a TP. Predicted changes
resembled those observed. FG Sge is a second candidate
for a late TP with presently characteristics of a
R~CrB. Despite these correspondences between observations and
predictions of the FF scenario, identification of majority
R~CrBs and EHes as fruits of the scenario is highly questionable.
On the other hand, the FF scenario
accounts well for the H-poor central stars of planetary
nebulae and PG1159 stars with their mix of He $\sim$ C $\sim$ O.

The DD scenario  exploits the fact
that a small fraction of binary
systems evolve to a pair of degenerate white dwarfs, for example,
a He white dwarf orbiting a C-O white dwarf. 
Angular momentum losses
via gravitational wave radiation or a magnetic field interaction ultimately cause
the merger of the stars: the He white dwarf is consumed by the C-O
white dwarf.
The merged star expands 
to red giant dimensions. 
This idea was put forward by Webbink (1984), and Iben \&
Tutukov (1985). Iben, Tutukov, \& Yungelson (1996) showed not only that this
binary model can account for the observed number of R~CrBs and EHes in the Galaxy
but mention several different production channels that might account for differences
of composition.
In this scenario, the coexistence of high abundances
 of C and N at the surface of
\rcb stars is explained by invoking  mixing between the C-O WD which
contains C but no N, and the He WD which contains N but no C. The mixing
takes place possibly at the time of merging (Iben \& Tutukov 1985).
Asplund et al. (2000) speculate that
the $rp$-process,  primarily $\alpha$-particle capture,
 might synthesize the intermediate mass elements Na -- S, and so account for
the unusual Si/Fe and S/Fe ratios of minority \rcb stars.
Detailed calculations of the nucleosynthesis expected from a merger
have not been reported. Saio \& Jeffery (2000) have examined the evolution
of two merged He white dwarfs and could explain the observed properties of
low-luminosity (or high-gravity) EHes, in particular of V652\,Her, with
its low C/He ratio.

An important fact about the compositions of the EHes and the R~CrBs
is that the O abundances are on average the `normal' values (i.e., [O/M]
is approximately centred on 0), and the maximum N abundances are consistent
with conversion of  the initial C and O, as judged by the metallicity
[M] (derived from the mean of Si and S), to N by the CNO-cycles.
In the DD scenario, these
facts must emerge from details of the merging process.
The following simple picture of a merger of a He white dwarf with a
C-O white dwarf
attempts to illustrate this point.

The He white dwarf contains primarily He with N, the principal
product of H-burning, as the second most
abundant element but is not enriched in $s$-process elements. 
A thin skin of H completes the structure. The C-O white dwarf
may contain a thin shell which was part of the He-shell of the
AGB star. This shell will contain He, C, and O, and is likely
enriched in the $s$-process heavy elements. Below this shell, the bulk of
the white dwarf will be C and O. It is worth noting that this scenario
likely accounts for the absence of $s$-process enrichment in the
EHes and R CrBs; the He white dwarf will not have been exposed
to a neutron source, and while the surface of the C-O white
dwarf may be $s$-process enriched having been the core of an AGB star,
little material from this star is needed to account for the C/He ratio
of our stars.
Merging may resemble the mixing of our simple recipe. Nitrogen
is provided by the accreted material from the He white dwarf,
and provided that this material dominates the atmosphere of the
merged star, the N abundance will approach that expected from
conversion of initial C and O to N. The C/He ratio is
determined by the mixing of the accreted material (He) with
the outer layers of the C-O white dwarf (a mixture of He, C, and O).
With mass fractions of unity for He in the accreted material and,
perhaps, about 0.2 for the C-O material, it is clear that  only if the
accreted material dominates the merged star can the observed low C/He ratios
be obtained. Similarly, a low O/M ratio is achieved only if the
accreted material dominates. Our rough calculation suggests that
a 10 to 1 mix of the He white dwarf to the C-O white dwarf
produces a merged star with a composition similar to the observed
compositions. Theoretical calculations have yet to determine the
mixing fraction, and the range that it may assume.
Our assumption of quiescent merging is possibly incorrect. Accretion
of material may ignite a carbon-oxygen shell. Thermal flashes
and mixing may ensue. Such flashes in the C-O white dwarf were considered by
Saio \& Nomoto (1998), and by Saio \& Jeffery (2000)
considering accretion of He-rich material by a He white
dwarf. The presence of freshly synthesized material in the merged star
enhances the coincidence that the composition matches the `normal'
O/M of the observed stars.

In Saio \& Jeffery's (2000) calculations, the accreted
He-rich material was  presumed to come from a second white dwarf in a
merger. This model was devised to account for V652\,Her, the hot
pulsating EHe with the low C/He ratio. Emphasis was placed not on matching
the observed chemical composition but on fitting the pulsations
and their time derivative. He-shell flashes occur in the
star allowing for some C to be mixed to the surface following
the first flash.
Since V652\,Her has the composition of a He-rich layer that results from
CNO-cycling, the merged star replicates the observed composition.
Constraints on mergers of He with C-O white dwarfs are more
severe with some elements provided primarily by the accreted
material, others by the C-O white dwarf, with some elements
dependent on the nucleosynthesis in the flashes and the mixing
between the flashed material and the envelope;
calculations are awaited with interest.

In summary, the FF scenario cannot account for the
typical EHe and \rcb star. The DD scenario remains a 
possibile explanation, but, perhaps, only until detailed calculations are reported!
Existence of minority \rcb stars with
spectacular Si/Fe and S/Fe ratios may be an indication of yet
another scenario. 

\section{Conclusions}

Our sample of cool EHes shows that their compositions resemble 
in many ways those of hot EHes and the R~CrBs; they might
be termed transition objects between these two groups of
He-rich stars. 
Close inspection of the analysed EHes - hot and
cool - and the R~CrBs - majority and minority - show a significant
difference in the N abundances. 
If the initial metallicity of a star
is assessed from its present Si and S abundances, the N abundance
of majority R~CrBs implies full conversion of initial C and O to N
by the CNO-cycles.
EHes show  generally lower N abundances
with several exhibiting only mild N enrichment.
This mildness appears misleading because N provided by conversion of 
the initial C and O to N may have been processed by $\alpha$-captures
to $^{22}$Ne; there is evidence for this neon in several stars.
 Their different N abundances could suggest that
the R~CrBs and EHes are not
on the same evolutionary path although their formation may be triggered by
similar events. Alternatively, the higher N abundance of the R CrBs
resulted when EHes evolved to red giants and a little C was
processed by the residual H to N. 

A majority of the stars show evidence of three components in their
atmosphere: a residue of normal H-rich material, substantial
amounts of H-poor CN(O)-cycled material, and C- (and O-)  rich 
material from gas exposed to He-burning. This combination could
be provided by a single star, as in the FF scenario, or by a merger
of white dwarfs, as in the DD scenario. Although the FF scenario
accounts for Sakurai's object and other stars
(e.g., the H-poor central stars of planetary nebulae),
present simulations imply much higher
C/He and O/He ratios than are observed in EHes and R~CrBs. 
The observed abundances imply constraints on the
merger process and subsequent evolution of the merged He-rich
star that results from the DD scenario and merger of a He white
dwarf with either a C-O or a He white dwarf. Theoretical evaluations
of the DD scenario have yet to grapple with these constraints. 
There remains too the intriguing problem of the minority R~CrBs
and similar stars with extraordinary ratios of Si/Fe and S/Fe, for example.
Is there a nuclear origin for these anomalies? Or are they
the result of a winnowing of dust from gas? 

Although an  observer's lament of `more observations' applies - for
example, measurements of Ne and Mg in R~CrBs are of interest,
also the $s$-process elements in EHes are crucial to see whether
$^{22}$Ne($\alpha$,n)$^{25}$Mg can be a neutron source - the louder
plea might be for additional theoretical work on the FF and,
especially, on the DD scenario. Additionally, it should not be
assumed that the origins of EHes and R~CrBs are to
be found in the DD or FF scenarios. 

Finally, it is surely of interest that H in these He-rich stars with the
exception of the two EHe stars with a low C/He ratio is depleted by at
least 3 dex. This raises the question - Why are there no mildly H-poor
relatives of the EHes in the same temperature range? To which  an answer may be -- there are but they are
difficult to identify. In support of this notion, we note that, in the
case of the R~CrBs that are more easily betrayed by their 
distinctive declines, H has been detected with an abundance 2 dex greater 
than in the most H-rich EHe where  H is depleted by just a factor
of 1.3 dex. An alternative answer may be that conversion of a H-rich to
a He-rich star is almost always so complete that mildly H-poor stars
do not exist, or exist for only a very short time. Discovery of
mildly H-poor relatives of the EHes and also additional R~CrBs
would be of great interest. Certainly, at effective
temperatures of the EHes and hotter, there are stars of intermediate H-deficiency. At the highest
temperatures, the H abundance is difficult to determine from optical
spectra. One may expect that mild cases of H deficiency arise for
reasons unrelated to the origins of the EHes and R CrBs.

\section{Acknowledgements}
We thank John Lattanzio for helpful conversations.
This research was supported by the Robert A. Welch Foundation, Texas, and the
National Science Foundation (grant AST 9618414).

\appendix

\section{Error Analysis}

    The major sources of error in deriving the abundances are the line-to-line
    scatter of the abundances and uncertainty in the adopted stellar
    parameters. The stellar parameters
    are accurate to typically: $\Delta$\Teff~= $\pm$300 K,
    $\Delta$log $g$ = $\pm$0.5 [cgs] and $\Delta$$\xi$ = $\pm$1 km $s^{-1}$.
    The derivatives of mean ion abundances with respect to \Teff~and log $g$
    were calculated. The maximum error in each
    ion abundance is estimated. For the majority of the ions, these errors are smaller than 
the line-to-line scatter (standard deviation due to several lines belonging to the same ion). 
The mean maximum error in the derived abundances corresponding to the uncertainty 
in \Teff~and log $g$ is about 0.1 dex; the errors due to the uncertainty in 
$\xi$ are negligible when compared to that due to uncertainties in the other parameters.

     The mean abundance of an element $X$, was calculated as;\\

$<X> = \frac{w_1 <X I> + w_2 <X II> + ...}{w_1 + w_2 + ...}$\\

\noindent
where $<X I>$, $<X II>$, ... are the abundance of element $X$
derived from neutral, singly ionized, ... lines of element $X$.
$w_1$, $w_2$, ... are weights i.e., \\

$w_1 = \frac{1}{(\delta (X I))^2}$\\

\noindent
$\delta (X I)$ is the error in $<X I>$ due to the uncertainty in
$T_{\rm eff}$ and log $g$.

The error in the ratio say $X/Fe$ was calculated as;\\

$\delta (X/Fe) = \frac{(\delta <X>) <Fe> - (\delta <Fe>) <X>}{<Fe>^2}$\\

\noindent
where \\

$(\delta <X>)^2 = \frac{1}{w_1 + w_2 + ...}$\\

\noindent
We take the modulus of \\

$\frac{(\delta <X>) <Fe> - (\delta <Fe>) <X>}{<Fe>^2}$\\

\noindent
to estimate the error in $X/Fe$.

   Abundance ratios are generally less affected by these uncertainties because most
    elements are sensitive to the stellar parameters in the same way.

The final abundances derived from neutral and ionized species of an element are separately given in Table 2
    for our programme stars. The hydrogen abundance is determined from the measured equivalent width of the  H$\alpha$
line.

\label{lastpage}

\end{document}